\RequirePackage{lineno} 
\documentclass[a4paper,10pt]{article}

\usepackage{amsmath}
\usepackage{amsthm}
\usepackage{amsfonts}
\usepackage{hyperref}
\usepackage{color}
\usepackage{multirow}
\usepackage[dvips]{graphicx}
\usepackage{subfigure}
\usepackage{amssymb}
\usepackage{epsfig}

\usepackage{setspace}

\usepackage[numbers, sort&compress]{natbib}

\usepackage[normalem]{ulem}

\newcommand{\comment}[1]{}
\numberwithin{equation}{section}

\begin{document}



\bibliographystyle{apsrev}

\title{\textbf{Towards the simplest simulation of incompressible viscous flows inspired by the lattice Boltzmann method}}


\author{$\textbf{Jun-Jie Huang}^{1, 2\footnote{Corresponding author. E-mail: jjhuang1980@gmail.com; jjhuang@cqu.edu.cn.}}$\\
	\\
	$ ^1$  Department of Engineering Mechanics, College of Aerospace Engineering, \\
	Chongqing University, Chongqing 400044, China \\
	$ ^2$ Chongqing Key Laboratory of Heterogeneous Material Mechanics, \\
	Chongqing University, Chongqing 400044, China}

\maketitle

\textbf{Abstract}

The lattice Boltzmann method (LBM) 
has gained increasing popularity in incompressible viscous flow simulations, but it uses many more variables  
than necessary. This defect was overcome by a recent 
approach that solves the \emph{more actual macroscopic equations} obtained through Taylor series expansion analysis of the lattice Boltzmann equations [Lu et al., J. Comp. Phys., 415, 109546 (2020)]. 
The key is to keep some small additional terms (SATs) to stabilize the numerical solution of the weakly compressible Navier-Stokes equations.  
However, there are many SATs that complicate the implementation of their method.
Based on some analyses and numerous tests, 
we ultimately pinpoint two essential ingredients for stable simulations:
(1) suitable density (pressure) 
diffusion added to the continuity equation; 
(2) proper numerical dissipation related to the velocity divergence  
added to the momentum equations. 
Then, we propose a simplified method 
that is not only easier to implement but noticeably faster
than the original method and the LBM. 
It contains much simpler SATs that \emph{only} involve the 
density (pressure) derivatives 
and it requires no intermediate steps or variables.
Besides, it is extended for two-phase flows with uniform density and viscosity.
Several test cases, including some two-phase problems under two dimensional, axisymmetric and three dimensional geometries, are presented to demonstrate its capability.
This work may help pave the way for the simplest simulation of incompressible viscous flows on collocated grids based on the artificial compressibility methodology.


\section{Introduction}\label{sec:intro}

An important and essential step in the numerical simulation of incompressible viscous flows is to find the pressure field, often by the solution of a Poisson equation~\cite{cmfd99book}.
It is well known that this step is rather time consuming and makes parallel computing more difficult. 
Over the past few decades, the lattice Boltzmann method (LBM) has become popular for incompressible flow simulation as it does not need to solve the Poisson equation~\cite{Chen1998}. LBM may be viewed as one type of artificial compressibility method (ACM)~\cite{he:lbm-ac, jcp10-acm, cma11-ac-lbm}
because the flow simulated by LBM is actually weakly compressible and the incompressibility condition is just approximately satisfied when the Mach number (Ma)  is low enough. 
Due to its explicit nature, LBM is relatively easy to implement and parallelize.
However, because of its kinetic origin, LBM has to use many particle distribution functions (PDFs), much more than the number of macroscopic variables, thus consuming more memory resources.
From certain perspective, LBM can be considered as a special finite difference method to solve the incompressible Navier-Stokes equations (NSEs)~\cite{jsc00-junk-klar, junk01fd-interp-lbm}.
However, the analyses in~\cite{jsc00-junk-klar, junk01fd-interp-lbm} involved the moments of the PDFs and the scheme proposed in~\cite{jsc00-junk-klar} used semi-implicit temporal discretization (still need to solve an elliptic problem). Is it possible to circumvent the PDFs and the related moments and directly evolve the weakly compressible NSEs explicitly?
The answer has been given recently in~\cite{jcp2020mame} which proposed an alternative method constructed upon the \emph{more actual macroscopic equations} (MAMEs) found by Taylor series expansion analysis of the lattice Boltzmann equations (LBEs). 
It was shown that direct discretization of the weakly compressible NSEs obtained by the usual Chapman-Enskog analysis  
are not stable;
in contrast, the MAMEs contain small additional terms (SATs) pivotal to stabilize the simulation~\cite{jcp2020mame}.
The SATs in the MAMEs are rather complex because they contain many terms involving the density (pressure), different velocity components and their derivatives in both space and time.
The numerical solution of the MAMEs introduces intermediate variables and uses a predictor-corrector procedure. Besides, proper boundary conditions must be supplied for some additional derivatives (which are not present in the original NSEs).
One may be curious on whether the MAMEs can be simplified with some nonessential terms discarded.
In this paper, we demonstrate that this can indeed be realized.

By making full use of the low-$Ma$ characteristics of the LBM, we further simplify the temporally discretized equations by discarding some terms of $O(Ma^{2})$ and keeping only derivatives of the density (pressure). Our method is based on the observation that the additional terms are already small and vanish as the time step $\delta_{t} \rightarrow 0$ (that is why the MAMEs can converge to the incompressible NSEs). The essential role of the SATs is to stabilize the computation.
There may be some room to adjust them as long as their magnitude is maintained at the same order without compromising the simulation (i.e., the stabilizing effects are still kept). From the design point of view,
one tends to make them as simple as possible and this can be optimally achieved by leaving \emph{only one scalar variable}. For incompressible flows, the pressure is the only choice (note that in ACM the pressure is tied to the density through an equation of state). Numerical tests indeed showed that under many situations it suffices to just keep the pressure terms.
Such simplifications make the method much easier to implement: the predictor and corrector steps in~\cite{jcp2020mame} are now combined into one single step and the intermediate variables are no longer necessary. The issue of boundary conditions for additional derivatives is resolved at the same time. As a result, the simulation needs even less memory and runs even faster.
The proposed simplified method is verified through several canonical tests for single-phase flows.
What is more, it effectiveness is also proven for two-phase flows with uniform density and viscosity (coupled with a phase field modeling of the interface dynamics).

In addition to the LBM, there are some other approaches for incompressible flow simulations based on the idea of AC, for instance, 
the kinetically reduced local Navier-Stokes (KRLNS)~\cite{prl05-krlns, pre07krlns},
the ACM with added dissipation~\cite{jcp10-acm} (denoted as \emph{suppressing checkerboard instability} (SCI) in~\cite{jcp18-gpe}), 
the link-wise ACM~\cite{jcp12-lwacm}, 
the methods of the entropically damped form of artificial compressibility (EDAC)~\cite{pre13-edac},
and the general pressure equation (GPE)~\cite{jcp18-gpe}.
The KRLNS uses a grand potential in the governing equations with an important term missing. It was later corrected by the EDAC which abandons the grand potential and uses common thermodynamic variables~\cite{pre13-edac}.
The governing equations in the EDAC method resemble the compressible NSEs except that the continuity equation for the density is replaced by an evolution equation for the pressure containing dissipative terms and the bulk viscosity component is neglected in the viscous stress tensor~\cite{pre13-edac}.
In actual implementation, the EDAC method usually employs a \emph{collocated} grid and various schemes can be used for spatial and temporal discretizations, for instance, the second order MacCormack scheme using a predictor-corrector sequence and the second order central scheme with high-order Runge-Kutta (RK) schemes for time marching~\cite{pre13-edac, cf17-edac-ibm, cma18-edac-turb}.
The GPE method is quite similar to the EDAC method in terms of the governing equations, but it uses a \emph{staggered} grid. 
For time marching, the third-order 
RK schemes are commonly used for GPE-based simulations~\cite{jcp18-gpe, tfc2020-gpe-turb, jcp20-gpe-turb}. 
In general, its implementation is 
more complicated than those on collocated grids and the simulation speed is  
slower than other one-step or two-step methods.
The second order version of the ACM with added dissipation also uses a \emph{collocated} grid
and intermediate variables for the pressure and velocity~\cite{jcp10-acm}, 
thus it may be viewed as a two-step method.
The link-wise ACM resembles the LBM to a significant degree, but it may circumvent the use of the  PDFs~\cite{jcp12-lwacm}.
It also uses a \emph{collocated} grid and 
is a one-step method.
However, the optimized implementation of the link-wise ACM without any PDFs involves many formulas that are quite complicated, especially in 3D.
Within the general ACM framework, the present method seems to be the simplest and easiest to implement: it uses a collocated grid, the 
second order schemes to discretize the spatial derivatives and one-step time marching. 

This paper is organized as follows. Section \ref{sec:mames} first introduces the MAMEs, its relation with the LBEs, and then presents the simplified MAMEs and its implementation. Next, the extension to two-phase flows is briefly described.
Section \ref{sec:results} provides the study of several common validation cases, including both single-phase and two-phase problems, by the proposed method and compares the numerical results with other reference ones.
Section \ref{sec:conclusion} concludes this paper with some discussions on future work.

\section{LBEs, MAMEs and the Simplified Formulation}\label{sec:mames}

The standard LBEs using single relaxation time read,
\begin{equation}\label{eq:lbe}
f_i (\boldsymbol{x} + \boldsymbol{e}_{i} \delta_{t}, t + \delta_{t} ) - f _i (\boldsymbol{x} , t)
= - \frac{1}{ \tau _f} [f _i (\boldsymbol{x} , t) - f _i ^{eq} (\boldsymbol{x} , t)], 
\end{equation}
where $\boldsymbol{e}_{i} = c (k_{x} \boldsymbol{e}_{x} + k_{y} \boldsymbol{e}_{y} + k_{z} \boldsymbol{e}_{z})$ is the discrete velocity ($c$ is the magnitude of lattice velocity, $k_{x}$, $k_{y}$ and $k_{z}$ are non-negative integers), $f _i$ is the PDF along $\boldsymbol{e}_{i}$
and $f _i ^{eq}$ is the corresponding equilibrium PDF, $\delta_{t}$ is the time step 
(the grid size $\delta_{x} = c \delta_{t}$), 
and the dimensionless relaxation parameter $\tau _f$ is related to the kinematic viscosity as  
$\tau _f = 0.5 + \nu/ (c_{s}^{2} \delta_{t})$ 
with $c_{s}$ being the sound speed in LBM
(usually $c_{s} = c / \sqrt{3}$). 
The macroscopic variables, including the density $\rho$ and the momentum
$\rho  \boldsymbol{u}$, are found from the PDFs as $\rho = \sum_{i} f_{i}$ and
$\rho  \boldsymbol{u} = \sum_{i} f_{i} \boldsymbol{e}_{i}$.
The pressure is tied to the density as $ p = \rho c_{s}^{2}$. 
By applying the Chapman-Enskog expansion analysis, 
one can find that the LBEs recover the following macroscopic equations (up to the second order in the Knudsen number)~\cite{jcp2020mame},
\begin{equation}\label{eq:nse-continuity-from-lbe}
\partial_{t} \rho = - \partial_{\alpha} (\rho u_{\alpha})  ,
\end{equation}
\begin{equation}\label{eq:nse-momentum-from-lbe}
\partial_{t} (\rho u_{\alpha}) = - \partial_{\beta} (\rho u_{\alpha} u_{\beta}
+  p \delta_{\alpha \beta} )  + \nu \partial_{\beta } \partial_{\beta} (\rho u_{\alpha}) 
\underline{+ \frac{\nu}{c_{s}^{2}} \partial_{\beta} \partial_{t} (\rho u_{\alpha} u_{\beta} + p \delta_{\alpha \beta})
+ 2 \nu \partial_{\alpha} (\partial_{\gamma} (\rho u_{\gamma}))  }.
\end{equation}
If Taylor series expansion analysis is applied to the LBEs, the equations to update the density and momentum read~\cite{jcp2020mame},
\begin{equation}\label{eq:mame-density}
 \rho (\boldsymbol{x}, t + \delta_{t})  =  \rho (\boldsymbol{x}, t)  - \delta_{t} \partial_{\alpha} (\rho u_{\alpha})
 \underline{+ \frac{1}{2} \delta_{t}^{2} \partial_{\alpha} \partial_{ \beta} (\rho u_{\alpha} u_{\beta} + p \delta_{\alpha \beta}) } + O(\delta_{t}^{3}) ,
\end{equation}
\begin{equation}\label{eq:mame-momentum}
\begin{split}
\rho u_{\alpha} (\boldsymbol{x}, t + \delta_{t})  &=  \rho u_{\alpha} (\boldsymbol{x}, t)  
- \delta_{t} \partial_{\beta} (\rho u_{\alpha} u_{\beta} + p \delta_{\alpha \beta})  
 + \nu \delta_{t} \partial_{\beta} \partial_{ \beta} (\rho u_{\alpha})  \\
&\underline{+ \bigg(\frac{\nu}{c_{s}^{2}} - 0.5 \delta_{t}\bigg) \delta_{t}  \partial_{t} \partial_{\beta} (\rho u_{\alpha} u_{\beta} + p \delta_{\alpha \beta}) 
	+ 2 \nu \delta_{t} \partial_{\alpha}   (\partial_{\gamma} (\rho u _{\gamma} ) )}
+ O(\delta_{t}^{3}) .
\end{split}
\end{equation}
These equations were claimed to be more actual and reasonable~\cite{jcp2020mame}. 
When compared with the weakly compressible NSEs, the underlined terms are SATs that help stabilize the simulation.
Note that eqs. \ref{eq:mame-density} and \ref{eq:mame-momentum} are partially discrete in time
and a predictor-corrector procedure was applied to handle the time derivative on the right hand side (RHS) of eq. \ref{eq:mame-momentum}. Specifically, the predictor step includes,
\begin{equation}\label{eq:mame-density-predictor}
\rho^* = \rho^n - \delta_{t} \partial_{\alpha} (\rho u_{\alpha})^{n}  
+  0.5 \delta_{t}^{2} \partial_{\alpha}\partial_{ \beta} (\rho u_{\alpha} u_{\beta} + p \delta_{\alpha \beta}) ^{n} ,
\end{equation}
\begin{equation}\label{eq:mame-momentum-predictor}
(\rho u_{\alpha})^{*} 
  =  (\rho u_{\alpha} )^{n}
- \delta_{t} \partial_{\beta} (\rho u_{\alpha} u_{\beta} + p \delta_{\alpha \beta})^{n}  
+ \nu \delta_{t} \partial_{\beta }\partial_{\beta} (\rho u_{\alpha})^{n}  
	+ 2 \nu \delta_{t} \partial_{\alpha}   (\partial_{\gamma} (\rho u _{\gamma} )^{n} ) ,
\end{equation}
and the corrector step includes,
\begin{equation}\label{eq:mame-density-corrector}
\rho^{n+1} = \rho^*  ,
\end{equation}
\begin{equation}\label{eq:mame-momentum-corrector}
(\rho u_{\alpha})^{n+1} 
=  (\rho u_{\alpha} )^{*}
+ \bigg(\frac{\nu}{c_{s}^{2}} - 0.5 \delta_{t}\bigg)   
[ \partial_{\beta} (\rho u_{\alpha} u_{\beta} + p \delta_{\alpha \beta})^{*} 
-\partial_{\beta} (\rho u_{\alpha} u_{\beta} + p \delta_{\alpha \beta})^{n}  ] .
\end{equation}
Since $\rho^{n+1} = \rho^* $ and $p = \rho c_{s}^{2}$, the density and pressure are not modified in the corrector step and only the velocity changes.
This predictor-corrector approach needs to compute and store the intermediate velocity $u_{\alpha} ^{*}$
and also has to calculate some derivatives of the intermediate variables (e.g., $\partial_{\beta} (\rho u_{\alpha} u_{\beta})^{*}$). 
It is noted that eqs. \ref{eq:nse-continuity-from-lbe} and \ref{eq:nse-momentum-from-lbe}
were unstable whereas eqs. \ref{eq:mame-density} and \ref{eq:mame-momentum} were stable 
when they are solved numerically by the same predictor-corrector procedure~\cite{jcp2020mame}.

In general, when one solves the incompressible 
NSEs, the additional terms are \emph{error terms that should be sufficiently small} so that they do not affect the accuracy of the numerical solutions.
Unlike the original terms in the NSEs 
(which have certain physical meanings and must be strictly followed in the numerical solution), 
the SATs have \emph{no real physical meanings} 
and their specific forms could possibly be adjusted. 
Of course, the adjustments must satisfy two requirements: 
(1) the magnitude of the additional terms is small enough (thus not altering the true solution) and converges towards zero as $\delta_{x} \rightarrow 0$ and $\delta_{t} \rightarrow 0$; 
(2) their stabilizing effect must be kept in the 
simulations with finite $\delta_{x}$ and $\delta_{t}$.
In LBM simulation of incompressible flows (assuming $c_{s}^{2} = c^{2} /3 $), one has 
$\partial_{\beta} (\rho u_{\alpha} u_{\beta} + p \delta_{\alpha \beta} ) = c^{2}  [\partial_{\beta}  (  \rho u_{\alpha} u_{\beta} / c^{2}) +  \partial_{\alpha} \rho / 3 ]$
where the two terms in the square brackets are both of $O(Ma^2)$.
When $\partial_{\beta} (\rho u_{\alpha} u_{\beta} + p \delta_{\alpha \beta} )$ appears in the SATs, it \emph{only serves to stabilize the computation and has no other roles}. 
Thus, provided that both terms are of the same order, one may keep only one of them without changing the order of magnitude of the SATs (as a whole). 
From these arguments, one may neglect $\partial_{\beta} (\rho u_{\alpha} u_{\beta})$ in the additional terms while satisfying the first requirement above.
It seems difficult 
to prove that the omittance of $\partial_{\beta} (\rho u_{\alpha} u_{\beta})$ also satisfies the second requirement.
Nevertheless, a number of tests, including both steady and unsteady, two dimensional (2D) and three dimensional (3D) problems, showed that the stablizing effect was indeed still there.
After neglecting $\partial_{\beta} (\rho u_{\alpha} u_{\beta})$, the correction in eq. \ref{eq:mame-momentum-corrector} becomes
$(1 / c_{s}^{2}) (\nu - 0.5 c_{s}^{2} \delta_{t})  \partial_{\alpha}  (p^{*} -p^{n}) =    (\nu - 0.5 c_{s}^{2} \delta_{t})  (\partial_{\alpha}  \rho^{*} -  \partial_{\alpha} \rho^{n})$. 
Besides, the additional term $2 \nu \delta_{t} \partial_{\alpha}   (\partial_{\gamma} (\rho u _{\gamma} ) )$ in eq. \ref{eq:mame-momentum-predictor} can be approximated as 
$ 2 \nu \delta_{t} \partial_{\alpha}   (- \partial_{t} \rho) = - 2 \nu \partial_{\alpha}   (\delta_{t}  \partial_{t} \rho)
\approx  - 2 \nu (\partial_{\alpha} \rho ^{n+1}  - \partial_{\alpha} \rho ^{n})$
by using the continuity equation. 
Using all these approximations and combining the predictor and corrector steps, we propose the following semi-discrete equations for the weakly compressible  
NSEs with $p = \rho c_{s}^{2}$,
\begin{equation}\label{eq:mame-density-simple}
\rho^{n+1} = \rho^n - \delta_{t} \partial_{\alpha} (\rho u_{\alpha})^{n}  
\underline{+  0.5 c_{s}^{2} \delta_{t}^{2} \partial_{\alpha \alpha}  \rho^{n}} ,
\end{equation}
\begin{equation}\label{eq:mame-momentum-simple}
(\rho u_{\alpha})^{n+1} 
=  (\rho u_{\alpha} )^{n}
- \delta_{t} \partial_{\beta} (\rho u_{\alpha} u_{\beta} + \rho c_{s}^{2} \delta_{\alpha \beta})^{n}  
+ \nu \delta_{t} \partial_{\beta}\partial_{ \beta} (\rho u_{\alpha})^{n}  
\underline{- (\nu + 0.5 c_{s}^{2} \delta_{t})   
( \partial_{\alpha} \rho^{n+1} -\partial_{\alpha}  \rho^{n}  ) }.
\end{equation}
The remaining spatial derivatives are discretized by the second order centered schemes, for example, in 2D
\begin{equation}\label{eq:centered-fd-drhodx-2d}
\partial_{x} \rho \vert_{i,j} = (\rho_{i+1,j} - \rho_{i-1,j} ) / (2 \delta_{x}),
\end{equation}
\begin{equation}\label{eq:centered-fd-laplacian-rho-2d}
 (\partial_{x} \partial_{x} \rho + \partial_{y} \partial_{y} \rho) \vert_{i,j} = (\rho_{i+1,j} + \rho_{i-1,j} + \rho_{i,j+1} + \rho_{i,j-1} - 4 \rho_{i,j}) / \delta_{x}^{2},
\end{equation}
The underlined terms in eqs. \ref{eq:mame-density-simple} and \ref{eq:mame-momentum-simple}
are the simplified SATs to stabilize the simulation, 
which \emph{only involve the derivatives of the density}. 
Since $\rho^{n+1}$ in eq.  \ref{eq:mame-momentum-simple} is directly found from eq. \ref{eq:mame-density-simple}, it is fully explicit and does not need intermediate variables.
It is noted that the SAT in eq. \ref{eq:mame-density-simple} resemble that in the pressure evolution equation in~\cite{jcp10-acm} (the coefficient before $\partial_{\alpha} \partial_{ \alpha}  \rho^{n}$ differs).
It was mentioned in~\cite{jcp10-acm} that such an additional term was added to overcome the checkerboard instability for the pressure.
We note that eq. \ref{eq:mame-density-simple} also resembles 
the pressure equation in the EDAC~\cite{pre13-edac} and
the GPE 
in~\cite{pla17-gpe, jcp18-gpe}, 
both of which contain an dissipation term proportional to the Laplacian of the pressure.  
The SATs in eq. \ref{eq:mame-momentum-simple} stabilize the simulation most likely in a way similar to the dissipation due to the bulk viscosity~\cite{dellar01pre, jcp12-lwacm} though it looks to be somewhat different. 
We argue that this particular form may be better as it only involves the density (pressure) gradient which is already calculated for eq. \ref{eq:mame-momentum-simple} (in contrast, the other forms require the gradient of velocity divergence).
When there is a body force $g_{\alpha}$ along the $x_{\alpha}-$direction, 
one only has to add a term $ \delta_{t} \rho^{n+1} g_{\alpha}$
on the RHS of eq. \ref{eq:mame-momentum-simple}.
Like the LBM, our method also uses the collocated arrangement of discrete variables in space.
Table \ref{tab:cmp-three-methods} compares the LBEs, the MAMEs~\cite{jcp2020mame} and the present SMAMEs (for Simplified MAMEs).

\begin{table}
	\begin{center} 
		\begin{tabular}{|c|c|c|c|} \hline
			Method & LBEs & MAMEs & SMAMEs 	\\\hline
			Variables (2D) & $\rho$, $u$, $v$, $f_{i}$ ($i = 0, 1, \cdots, 8$) & $\rho$, $u$, $v$, $u^{*}$, $v^{*}$ & $\rho$, $u$, $v$   \\\hline
			Number(\#) of variables (2D) & 12 & 5 & 3 \\\hline
			\# of evolution eqns (2D) & 9  (D2Q9)& 5 & 3 \\\hline
			\# of small stablizing terms (2D)& N.A. & $5+(2+6)\times 2 = 21$ & $2+2\times 2 = 6$ \\\hline
			Variables (3D) & $\rho$, $u$, $v$, $w$, $f_{i}$ ($i = 0, 1, \cdots, 18$)  & $\rho$, $u$, $v$, $w$, $u^{*}$, $v^{*}$, $w^{*}$  & $\rho$, $u$, $v$, $w$   \\\hline
			\# of variables (3D) & 23 & 7 & 4 \\\hline
			\# of evolution eqns (3D) & 19 (D3Q19) & 7 & 4 \\\hline
			\# of small stablizing terms (3D)& N.A. & $9+(3+8)\times 3 = 42$ & $3+2\times 3 = 9$ \\\hline
			\end{tabular}
	\end{center}
	\caption{Comparisons between the LBEs, the original MAMEs, and the simplified MAMEs for incompressible flow simulations (assuming $p = \rho c_{s}^{2}$). In the original MAMEs, the corrector step does not alter the density obtained in the predictor step (i.e., $\rho^{n+1} = \rho^{*}$) and they are regarded as one variable. For the LBEs, D2Q9 and D3Q19 velocity models are assumed for 2D and 3D cases, respectively. When counting the stabilizing terms, the Laplacian is treated separately (e.g., in 2D $\partial_{\alpha \alpha}  \rho = \partial_{x x}  \rho + \partial_{y y}  \rho$ has two terms). In the LBEs, no explicit stablizing terms are added.}
	\label{tab:cmp-three-methods}
\end{table}

In addition to the above simplifications, 
we also extend the proposed method to two-phase flows with uniform density and viscosity.
Both fluids have the same density $\rho_{0}$ and kinematic viscosity $\nu$. Another term 
$\delta_{t} F_{ST, \alpha} = - \delta_{t} \phi \partial_{\alpha} \mu$ is added 
on the RHS of eq. \ref{eq:mame-momentum-simple} to account for the surface tension effect.
Here $\phi$ is the order parameter, $\mu = 4 a \phi (\phi^{2} - 1) - \kappa \partial_{\alpha }\partial_{\alpha} \phi$ is the chemical potential and the evolution of $\phi$ is governed by the Cahn-Hilliard equation (CHE)~\cite{cahnhilliard58, jacqmin99jcp},
\begin{equation}\label{eq:che}
\partial_{t} \phi + u_{\alpha} \partial_{\alpha} \phi = M \partial_{\alpha }\partial_{\alpha} \mu .
\end{equation} 
Here $M$ is the mobility, and $a$ and $\kappa$ are two constants related to the surface tension $\sigma$ and interface width $W$ as $a = 3 \sigma /(4 W)$, $\kappa = 3 \sigma W / 8$.
On a solid wall with a unit normal vector $\boldsymbol{n}_{w}$ and contact angle $\theta$, 
the no flux condition $\partial_{n_{w}} \mu = 0$ is applied for the chemical potential
and the wetting boundary condition $\partial_{n_{w}} \phi = - (2/W) \cos \theta (1 - \phi_{w}^{2})$ 
is applied for the order parameter ($\phi_{w}$ is the order parameter at the wall)~\cite{ijnmf15-wbc-pfs}.
The spatial derivatives in eq. \ref{eq:che} are discretized by the second order isotropic schemes
(see~\cite{hybrid-mrt-lb-fd-axisym} for details)
and its time marching uses the second order RK scheme.
It should be noted that for two-phase flows the real fluid density $\rho_{0}$ is a constant (set to $1$) whereas $\rho = p / c_{s}^{2}$ has small variations around its initial value (also set to $1$).

\section{Results and Discussions}\label{sec:results}

In this section, we present the results of several test cases obtained by using the new method and make comparisons with those in the literature and by other methods 
under the same simulation settings (i.e., same $\delta_{x}$ and $\delta_{t}$). 
Note that the LBM simulations below use the D2Q9 and D3Q19 velocity models for 2D and 3D problems, respectively, 
and the multiple relaxation time (MRT)~\cite{pre2000-lbe-theory}
or the weighted MRT model~\cite{jcp17-wmrt-lbm-mfflow}
for the collision step.
Uniform mesh and time step are used in all problems. 
For each problem, a characteristic length $L$ and characteristic velocity $U_{c}$ are chosen.
The characteristic time $T_{c} = L/U_{c}$ is divided into $N_{t}$ segments ($\delta_{t} = T_{c}/N_{t}$) 
and the characteristic length $L$ is discretized into $N_{L}$ segments ($\delta_{x} = L/N_{L}$).

\subsection{Taylor-Green vortex in 2D}\label{ssec:tgv}

The first case is the Taylor-Green vortex in 2D. The analytical solutions for this problem 
in the domain $[-1, 1] \times [-1, 1]$ (i.e., the characteristic length $L = 1$) 
with a characteristic velocity $u_{0} $ (set to $1$) are given by~\cite{jcp2020mame},
\begin{subequations}
	\begin{equation}\label{eq:tgv-anal-sol-u}
u_{analytical} (x, y, t) = - u_{0} \cos (\pi x/L)  \sin (\pi y/L) \exp ( - 2 \pi^{2} t / Re),
\end{equation}
	\begin{equation}\label{eq:tgv-anal-sol-v}
v_{analytical}  (x, y, t) = u_{0} \sin (\pi x/L)  \cos (\pi y/L) \exp ( - 2 \pi^{2} t / Re),
\end{equation}
	\begin{equation}\label{eq:tgv-anal-sol-p}
p_{analytical}  (x, y, t) =  p_{0} - 0.25 \rho_{0} u_{0}^{2} [\cos (2 \pi x/L) + \cos (2 \pi y/L) ] \exp ( - 4 \pi^{2} u_{0} t / (Re L)) .
\end{equation}
\end{subequations}
where $Re  = u_{0} L / \nu$ is the Reynolds number, $\rho_{0}$ is the density (set to $1$), 
and $p_{0}$ is the reference pressure (set to $p_{0} = \rho_{0} c_{s}^{2}$).
This problem is periodic in both the $x-$ and $y-$directions.
The initial fields are set according to eqs. \ref{eq:tgv-anal-sol-u}, \ref{eq:tgv-anal-sol-v} and \ref{eq:tgv-anal-sol-p} with $t = 0$.
The case at $Re = 20$ (same as~\cite{jcp2020mame}) is studied.

\begin{figure}[htp]
	\centering
	\includegraphics[trim= 1mm 1mm 1mm 1mm, clip, scale = 0.75, angle = 0]{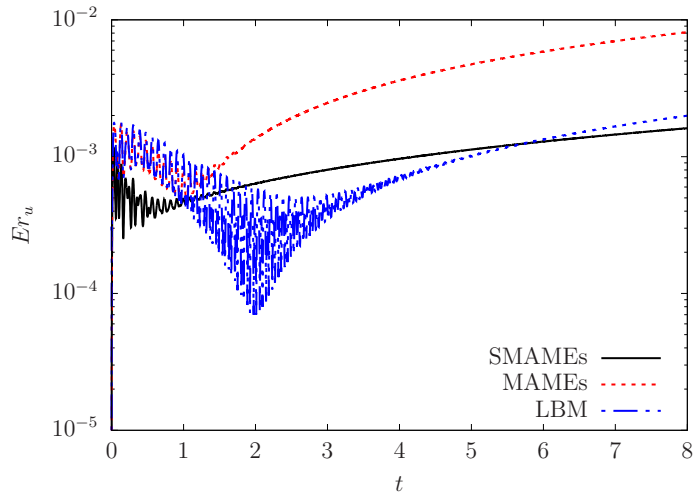}
	\caption{
		Evolutions of the error in the horizontal velocity component $Er_{u}$ in the simulation of the 2D Taylor-Green vortex.
		The Reynolds number is $Re = 20$ and the shared simulation parameters are
		$N_{L} = 50$ and $N_{t} = 1000$ ($c = 20$).
		The solid line is by the present method, the dashed line is by the original MAMEs and the dash-dot-dot line is by the LBM using MRT. 
	}
	\label{fig:cmp-dev-u-evol}
\end{figure}

We mainly focus on the error in the numerical solutions of the velocity component $u$ defined as follows, 
	\begin{equation}\label{eq:err-u}
Er_{u} (t)= \frac{ \sqrt{\sum_{i,j} (u_{numerical} (x_{i,j}, y_{i,j}, t) - u_{analytical} (x_{i,j}, y_{i,j}, t))^{2}} }{ \sqrt{\sum_{i,j} u^{2}_{analytical} (x_{i,j}, y_{i,j}, t) } } ,
\end{equation}
where the summation is for all nodes in the simulation domain.
Note the error $Er_{v} (t) $ is the same as $Er_{u} (t)$ due to the symmetry of this problem.
Figure \ref{fig:cmp-dev-u-evol} shows the evolutions of $Er_{u}$ over a relatively long period of time ($0 \leq t \leq 8$) by using the present method (SMAMEs), the original MAMEs in~\cite{jcp2020mame} and the LBM (using MRT~\cite{pre2000-lbe-theory}). 
For the LBM, the model parameters for MRT follow those in fig. 1 of~\cite{pre2000-lbe-theory}.
It is seen that for all three methods the deviations remain small (around $10^{-3}$) in the early stage ($t < 1$),
but after some time the deviations grow with time. At the end of simulation ($t=8$), the present method and LBM can still have reasonably good accuracy ($O(10^{-3})$) whereas the original MAMEs give less satisfactory results ($O(10^{-2})$).
With the same simulation parameters ($N_{L} = 50$ and $c = 20$), 
the computation times are $18.0$s, $23.3$s and $34.2$s 
for the present SMAMEs, the original MAMEs and the MRT-LBM, respectively.
That is, the present method saves about $25\%$ of the time compared to the original MAMEs
and it saves nearly one half of the time compared to the MRT-LBM.
At the same time, it is as accurate as the MRT-LBM and more accurate than the original MAMEs.
It is noted that the velocity magnitude at $t=8$ has decreased by three orders of magnitude (compared to its initial value) and it may be more difficult to closely follow the analytical solutions.

\begin{figure}[htp]
	\centering
	\includegraphics[trim= 1mm 1mm 1mm 1mm, clip, scale = 0.75, angle = 0]{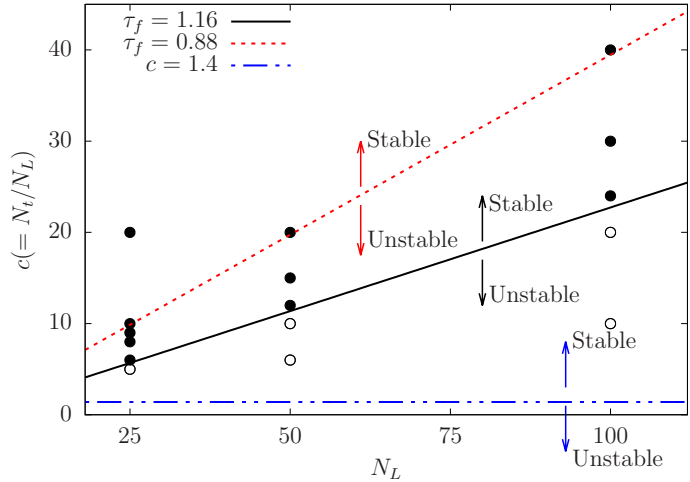}
	\caption{
		Stable and unstable regions in the $N_{L} - c$ plane for different methods to simulate the Taylor-Green vortex problem at $Re=20$. Note the grid size $\delta_{x} = 1/N_{L}$ and the time step $\delta_{t} = 1 / N_{t} = 1 / (c N_{L})$.
		The three lines approximately represent the critical conditions for the present method (solid), the original MAMEs (dashed) and the LBM using MRT (dash-dot-dot). 
		The stable regions are above the respective lines and the unstable regions are below them.
		The filled circles represent the stable cases and the empty circles represent the unstable cases for the present method (for clarity, the specific data for the other two methods are not plotted).
		The tests were performed for $0 \leq t \leq 4$. In the stable runs the errors do not increase abruptly.
		For the LBM runs, the MRT parameters follow those in fig. 1 of~\cite{pre2000-lbe-theory} and the errors are quite large when $c$ is too small (even though the runs do not blow up).
	}
	\label{fig:tgv-stability-diagram-re20}
\end{figure}

Figure \ref{fig:tgv-stability-diagram-re20} shows the stability diagram in the $N_{L} - c$ plane for three methods (SMAMEs, MAMEs, and LBM) for the same case at $Re=20$. 
It is seen that the stable region of the present method 
is larger on this map than that of the original method using the MAMEs.
In other words, under the same grid size, the present method can use a larger time step than the original method in~\cite{jcp2020mame}.
The difference in the minimal $c$ (or the maximal $\delta_{t}$) for a stable computation between the two methods increases as $N_{L}$ increases (i.e., the grid size $\delta_{x}$ decreases).
Among the three methods being compared, the LBM is the most stable and the minimal $c$ (to keep the simulation stable) is almost  constant when the mesh is refined.
However, it does not mean that the LBM can give reliable results irrespective of the lattice velocity $c$.
In fact, the LBM for incompressible flow simulations should follow the diffusive scaling ($\delta_{x} \sim O(\epsilon)$ and $\delta_{t} \sim O(\epsilon^2)$)~\cite{jsc00-junk-klar, jcp04-lbm-error, jcp05-lbe-asymp}.
That means the lattice velocity $c$ should satisfy $c \sim O(1/\epsilon) \sim O(N_{L})$.
Thus, even though LBM remains stable when $c \sim O(1)$, it does not satisfy the requirement for the simulation of incompressible flows. 
Overall, in terms of the stability performance, 
the present method is in between the original MAMEs and the LBM 
according to fig. \ref{fig:tgv-stability-diagram-re20}.

\begin{figure}[htp]
	\centering
	\includegraphics[trim= 1mm 1mm 1mm 1mm, clip, scale = 0.65, angle = 0]{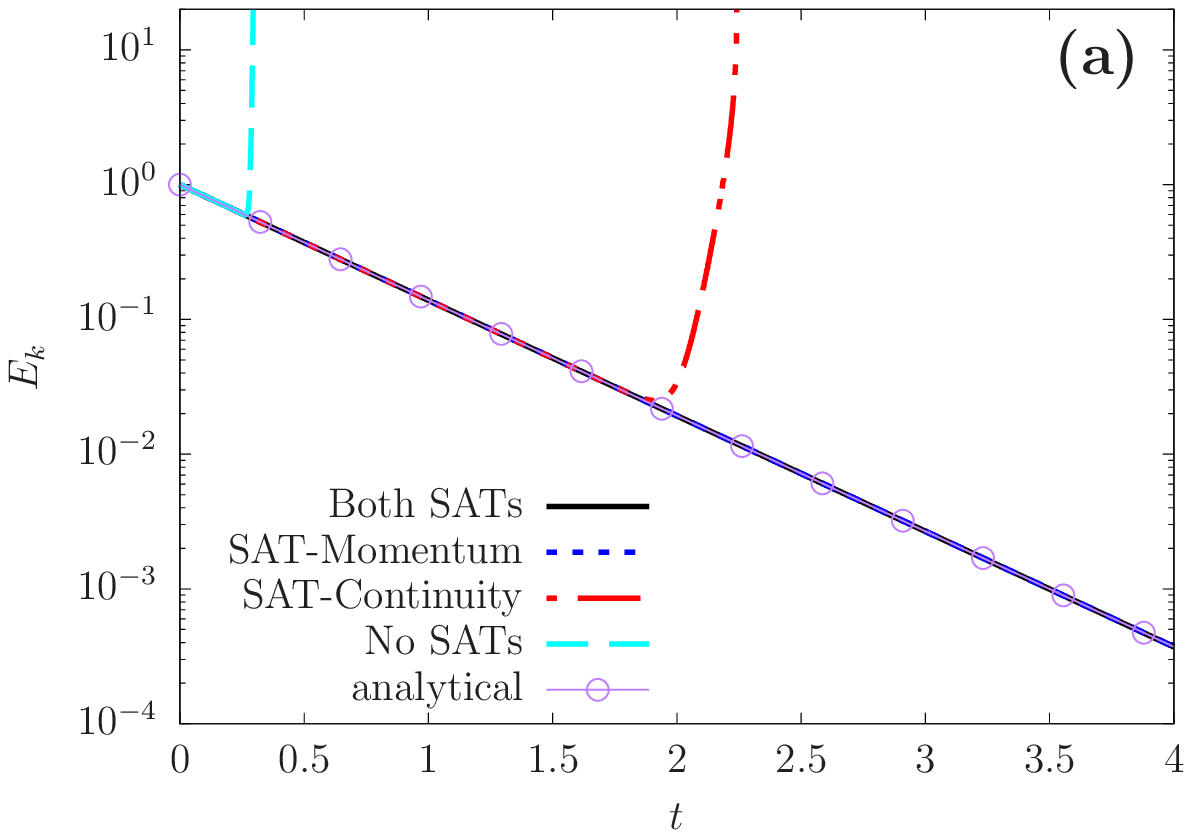}
	\includegraphics[trim= 1mm 1mm 1mm 1mm, clip, scale = 0.65, angle = 0]{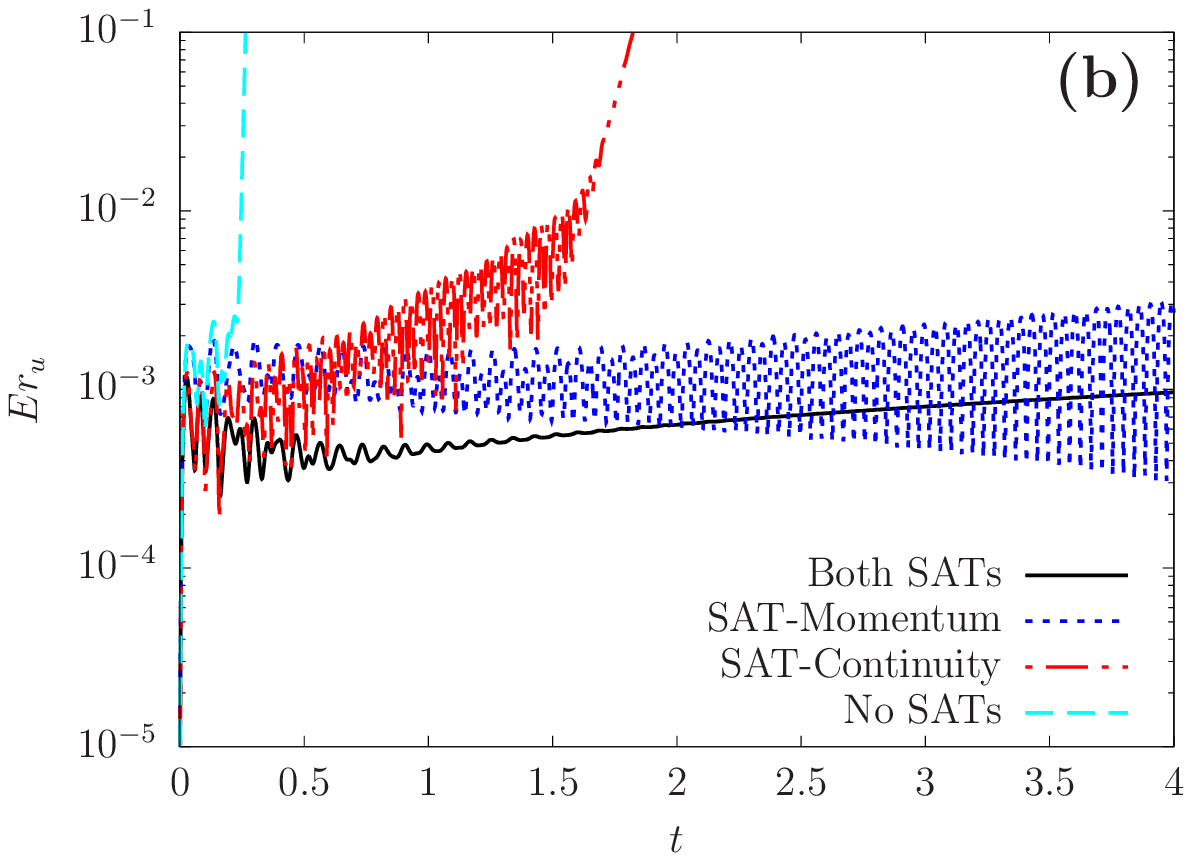}
	\caption{
		Evolutions of (a) the total kinetic energy $E_{k}$ 
		(b) the error in the horizontal velocity component $Er_{u}$.
		The solid line was obtained with both the SATs in eqs. \ref{eq:mame-density-simple} and \ref{eq:mame-momentum-simple},
		the dashed line was obtained with the SATs in the momentum equations (eq. \ref{eq:mame-momentum-simple}) only, 
		the dash-dot-dotted line was obtained with the SAT in the continuity equation (eq. \ref{eq:mame-momentum-simple}) only, 
		and the long-dashed line was obtained without any SAT.
		The shared simulation parameters are
		$N_{L} = 50$ and $N_{t} = 1000$ ($c = 20$).
	}
	\label{fig:cmp-ke-dev-u-evol-effect-add-term}
\end{figure}

Next, the effects of the SATs in eqs. \ref{eq:mame-density-simple} and \ref{eq:mame-momentum-simple}
are investigated. 
For this study, the total kinetic energy $E_{k} (t) = \int_{-1}^{1} \int_{-1}^{1} \frac{1}{2} [u^{2}(x,y, t)+v^{2}(x,y, t)] dx dy $ was also monitored.
From the analytical solution, one finds that $E_{k} (t)  = \exp (- 4 \pi^{2} t / Re)$.
Figure \ref{fig:cmp-ke-dev-u-evol-effect-add-term}
compares the evolutions of $E_{k}$ and $Er_{u}$ for a typical case at $Re=20$
as obtained by using four different simulation settings
(1) both the SATs in the continuity and momentum equations are included
(2) only the SATs in the momentum equations are included
(3) only the SATs in the continuity equation is included
(4) no SATs are included.
When the SATs are absent from eqs. \ref{eq:mame-density-simple} and \ref{eq:mame-momentum-simple}, both the kinetic energy and the error in $u$ quickly become very large.
When the SAT in the continuity equation is added, the situation improves slightly but the simulation still goes unstable ($E_{k}$ increases sharply) after some time. This indicates that the SATs in the momentum equations are crucial to maintain the stability.
With the SATs only added in the momentum equations, the computation remains stable and the total kinetic energy follows the analytical prediction, but the error in $u$ shows significant fluctuations. The reason may be that the SAT in the continuity equation, which is dissipative in nature, helps to damp out the oscillations in the density (pressure) field.
In contrast to the above three situations, 
when the SATs in the continuity and momentum equations are both added, 
the simulation is not only stable but also shows the least fluctuations in $Er_{u}$.
As noted earlier, the SATs in eqs. \ref{eq:mame-density-simple} and \ref{eq:mame-momentum-simple}
are much simplified compared with the original MAMEs in~\cite{jcp2020mame}.
Yet they are sufficient to keep the computation stable and provide accurate results.
This will be further demonstrated through other tests below.
It seems that it is difficult 
to further simplify the SATs.

\subsection{Lid-driven cavity flow in 2D and 3D}\label{ssec:driven-cavity}

\begin{figure}[htp]
	\centering
	\includegraphics[trim= 1mm 1mm 1mm 1mm, clip, scale = 0.6, angle = 0]{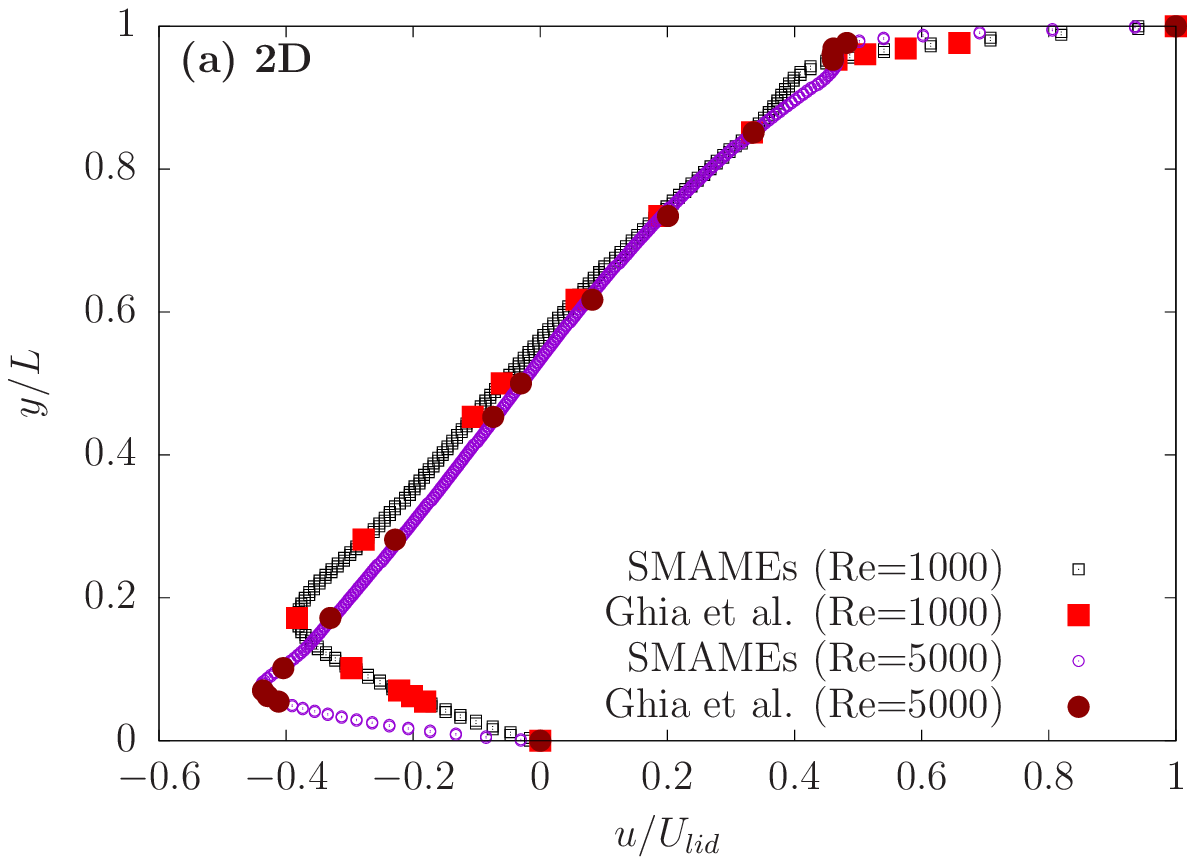}
	\includegraphics[trim= 1mm 1mm 1mm 1mm, clip, scale = 0.6, angle = 0]{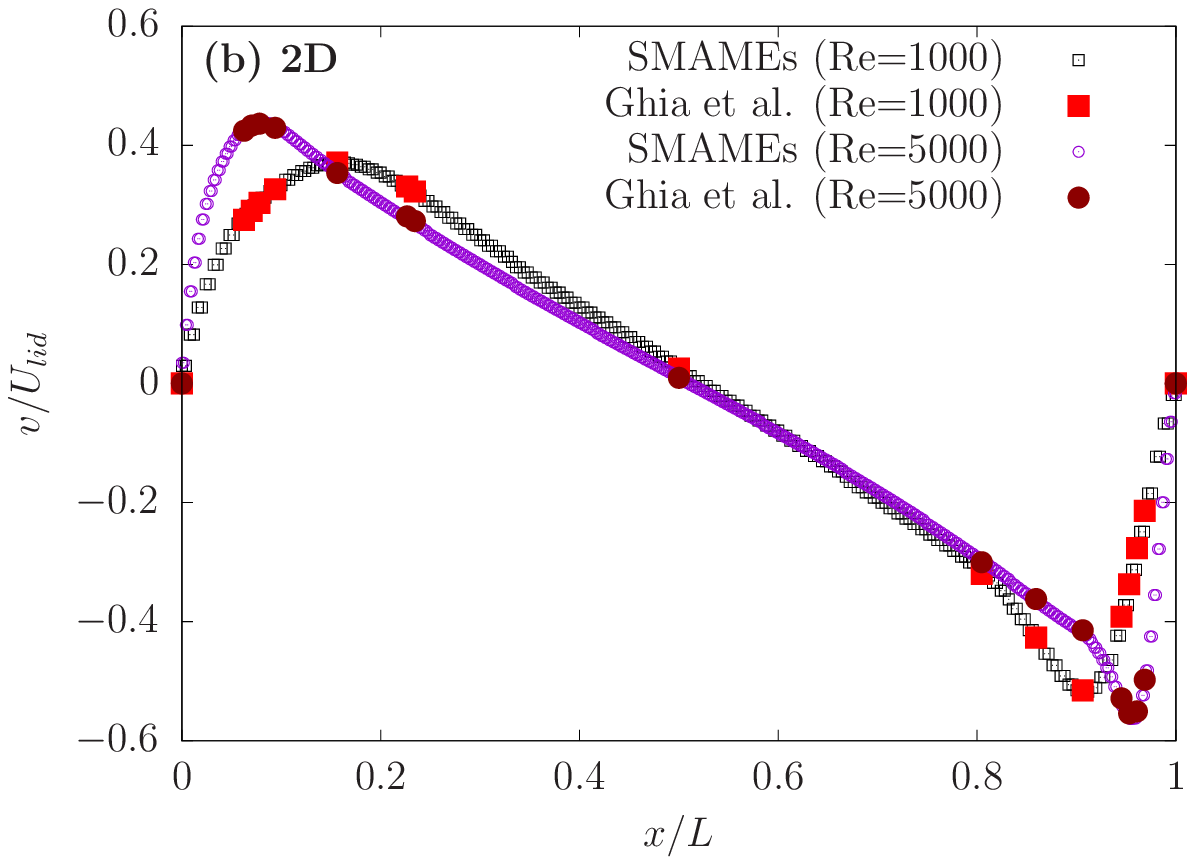}
	\includegraphics[trim= 1mm 1mm 1mm 1mm, clip, scale = 0.6, angle = 0]{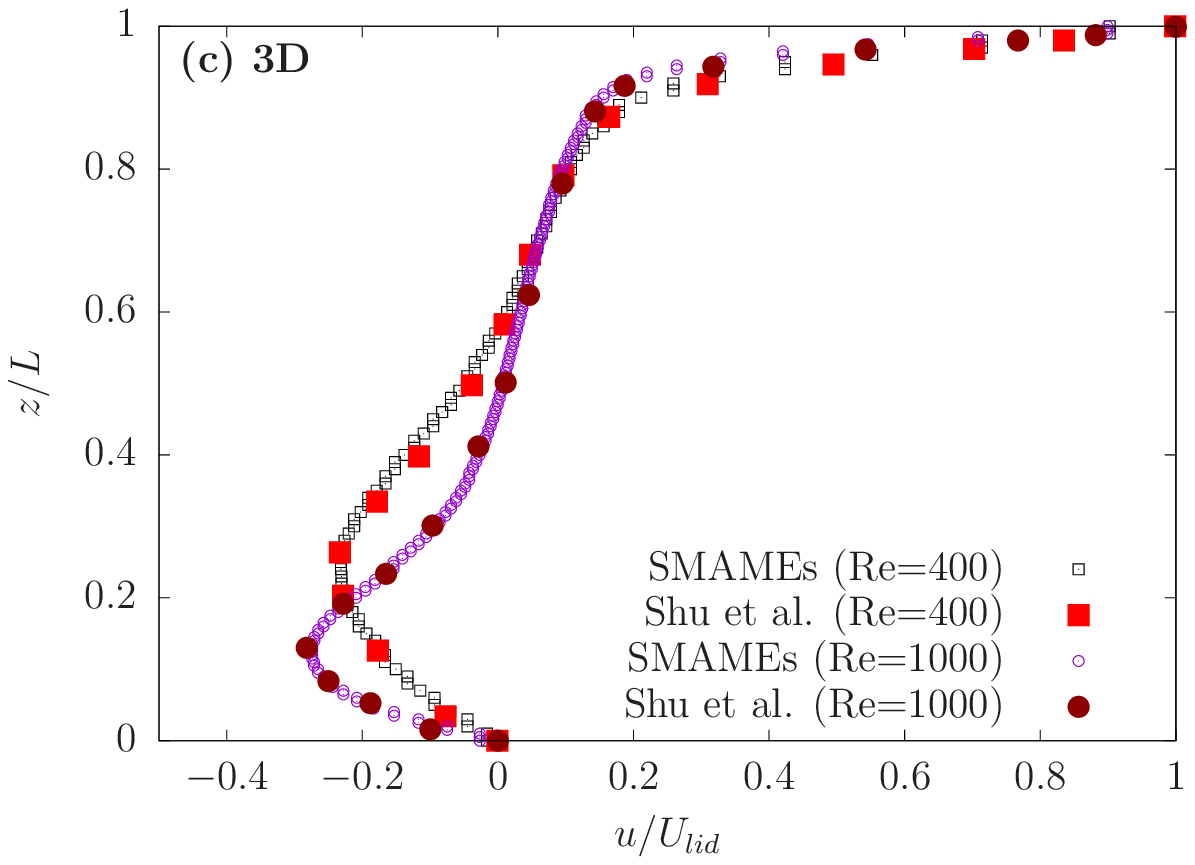}
	\includegraphics[trim= 1mm 1mm 1mm 1mm, clip, scale = 0.6, angle = 0]{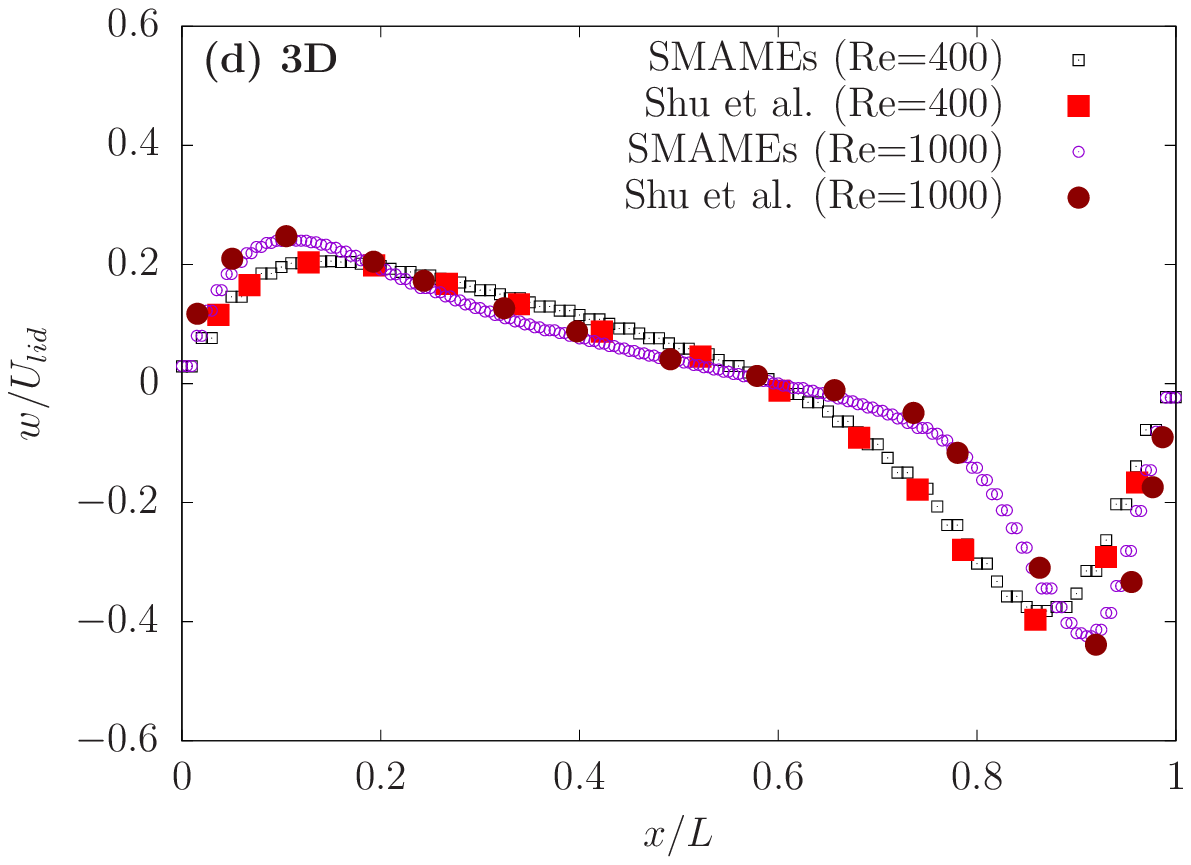}
	\caption{
		Velocity profiles along the centerlines 
		(a) $u(y)$ at $x=0.5$  and (b) $v(x)$ at $y = 0.5$ for the 2D cases;
		(c) $u(z)$ at $x=0.5$ and (d) $w(x)$ at $z=0.5$ in the plane $y=0.5$ for the 3D cases.
		The reference data for the 2D cases are from~\cite{jcp82-driven-cavity} 
		and those for the 3D cases are from~\cite{ijmpc03-tllbm3d}.
	}
	\label{fig:cmp-vp-driven-cavity}
\end{figure}

The second case is the lid-driven cavity flow. Both 2D and 3D situations were studied.
In 2D the domain is a square $[0, 1] \times [0, 1]$ enclosed by four solid walls (i.e., the side length $L$ is chosen as the characteristic length).
The top wall at $y = 1$ is moving with a constant velocity $(U_{lid},0) = (1,0)$ (i.e., the lid velocity is chosen to be the characteristic velocity). 
In 3D the domain is a cube $[0, 1]  \times [0, 1]\times [0, 1]$,
and the top wall at $z = 1$ is moving with a constant velocity $(U_{lid},0, 0) = (1,0, 0)$.
All other walls are stationary.
The Reynolds number is given by $Re = U_{lid} L / \nu$. 
The initial fields are set to be $u = v (= w) = 0$ and $\rho = 1$.
The criterion 
$\max (\vert \boldsymbol{u} \vert^{n+1} - \vert \boldsymbol{u} \vert^{n}) < 10^{-8}$ 
is used to determine whether the steady state is reached 
(i.e., the change in the velocity magnitude between two consecutive steps  
is less than $10^{-8} $ everywhere).
Several cases commonly used for benchmark studies were investigated, including $Re=100$, $400$,
$1000$ and $5000$ in 2D and $Re=100$, $400$ and $1000$ in 3D.
Here we only present the results of four cases at high $Re$ numbers ($Re=1000$ and $5000$ in 2D, and $Re=400$ and $1000$ in 3D).
For the 2D cases, the numerical parameters are $N_{L} = 128$, $c=20$ for $Re=1000$
and $N_{L} = 256$, $c=50$ for $Re=5000$.
For the 3D cases they are $N_{L} = 64$, $c=20$ for $Re=400$
and $N_{L} = 96$, $c=20$ for $Re=1000$.
Figure \ref{fig:cmp-vp-driven-cavity} gives the velocity profiles along selected centerlines
for the four cases. The data from~\cite{jcp82-driven-cavity, ijmpc03-tllbm3d} are also plotted for comparison.
The data from~\cite{jcp82-driven-cavity} were obtained by solving the incompressible NSEs using the vorticity-stream function formulation and have been widely used for benchmarking purposes.
The data from~\cite{ijmpc03-tllbm3d} were obtained by a special formulation of the LBM using nonuniform meshes.
It is seen that the present results are in good agreement with both reference results.
The results for other low $Re$ cases 
also agree well with the reference ones, but for conciseness they are not shown here.

\subsection{Doubly periodic shear layer}\label{ssec:dpsl}

\begin{figure}[htp]
	\centering
	\includegraphics[trim= 1mm 1mm 1mm 1mm, clip, scale = 0.65, angle = 0]{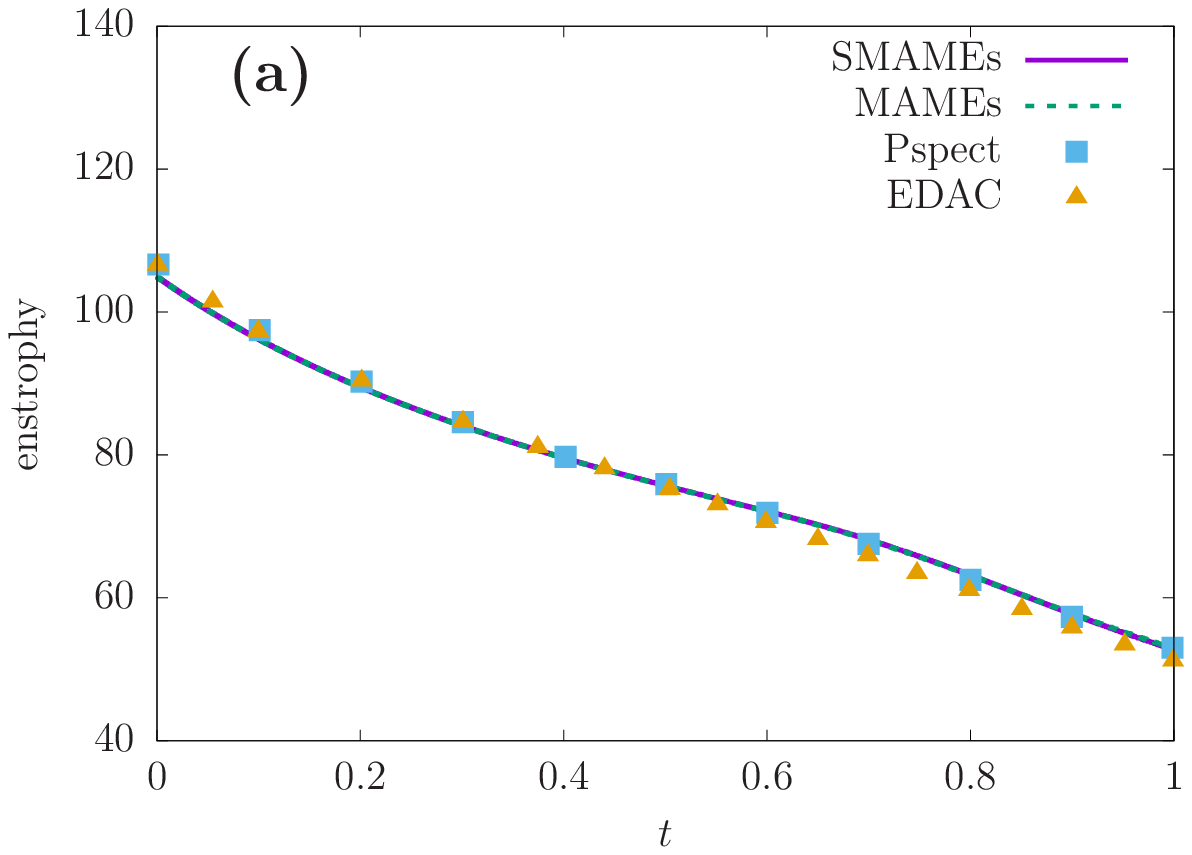}
	\includegraphics[trim= 1mm 1mm 1mm 1mm, clip, scale = 0.65, angle = 0]{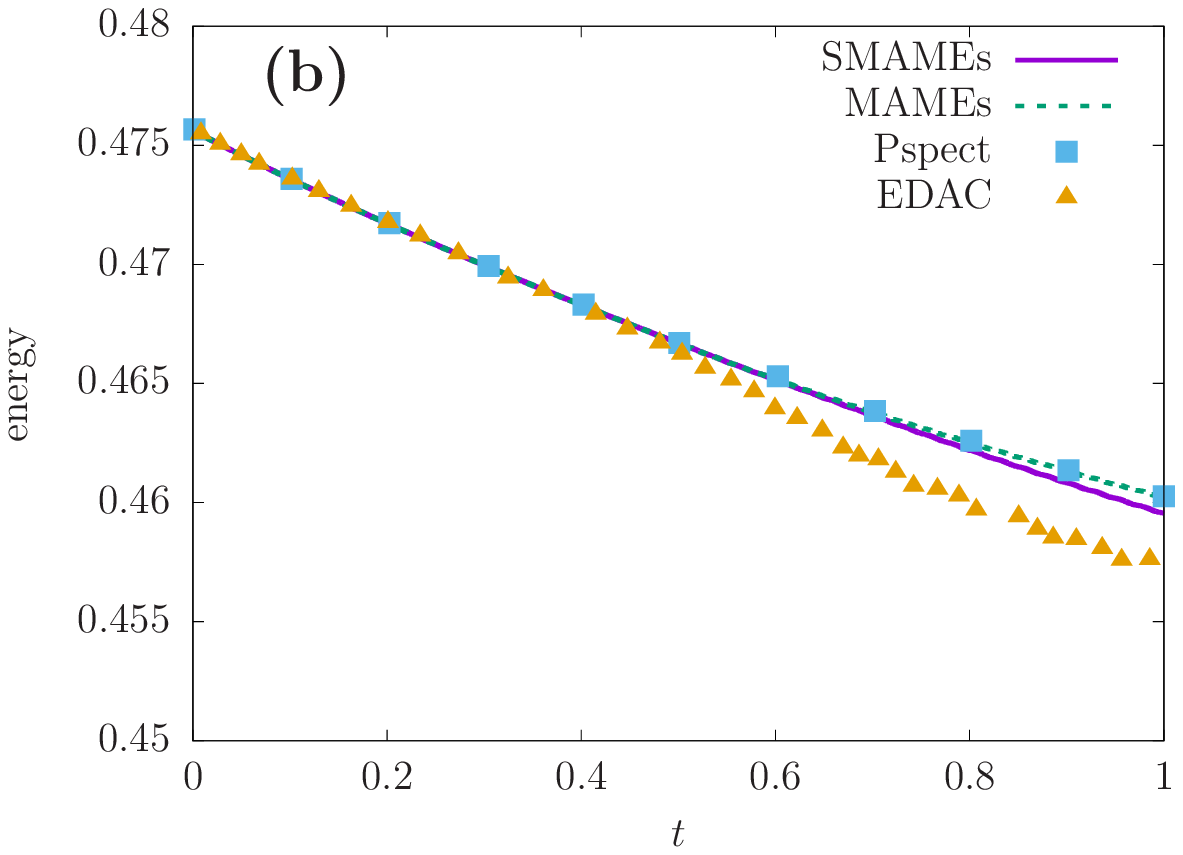}
	\caption{
		Evolutions of (a) the enstrophy (b) the energy for the doubly periodic shear layer
		at $Re = 10000$.
		The present simulation parameters are
		$N_{L} = 320$, $N_{t} = 16000$ ($c = 50$).
		For the EDAC results from~\cite{pre13-edac}, $N_{L} = 512$ and $N_{t} = 5120$ ($c=10$).
		Note that "Pspect" denotes the results obtained by a pseudospectral solve on a $768 \times 768$ grid
		~\cite{pre13-edac, jcp97-dpsl-test}.
	}
	\label{fig:cmp-dpsl-enstrophy-ke-evol}
\end{figure}

\begin{figure}[htp]
	\centering
	\includegraphics[trim= 1mm 1mm 1mm 1mm, clip, scale = 0.35, angle = 0]{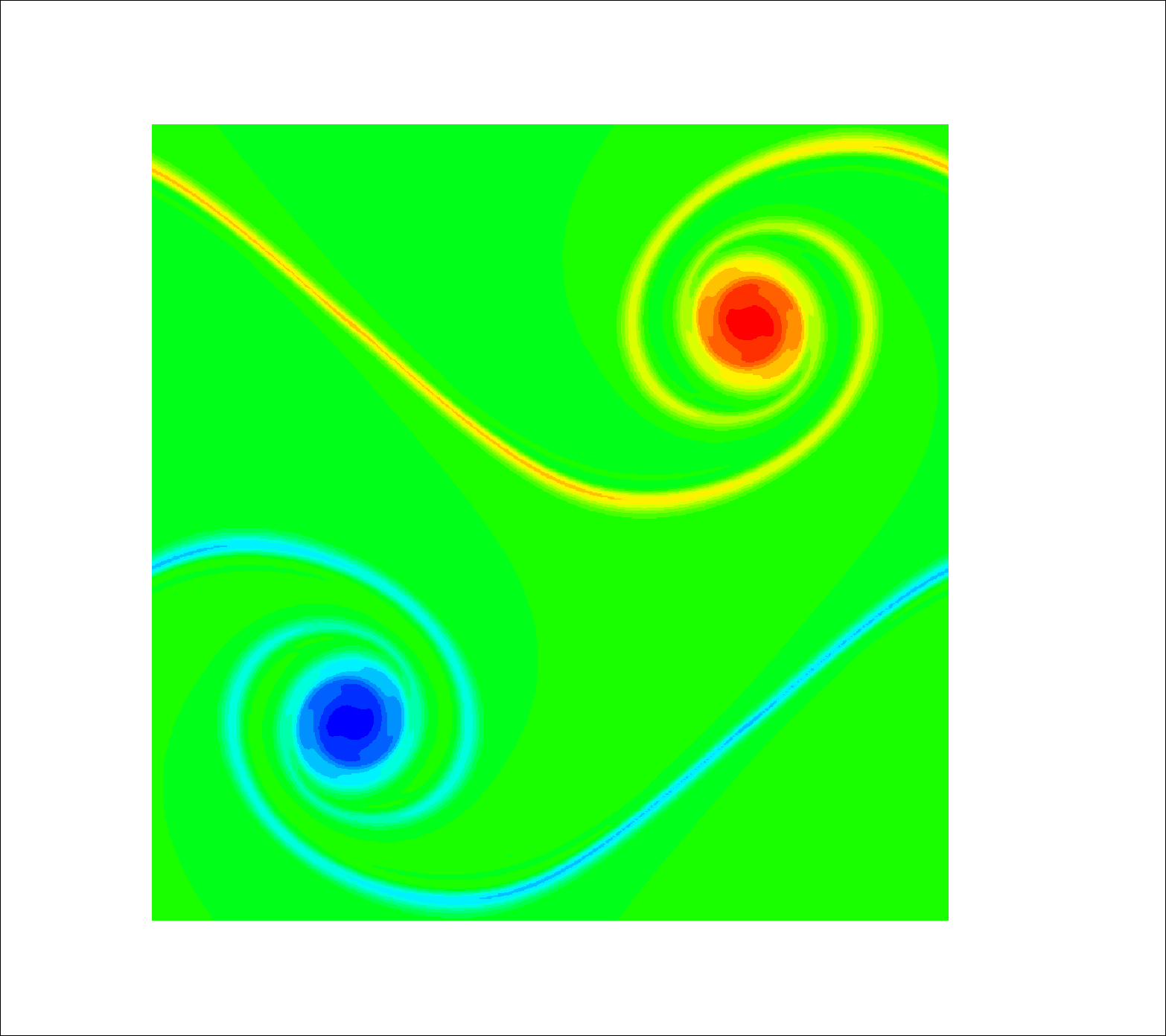}
	\includegraphics[trim= 1mm 1mm 1mm 1mm, clip, scale = 0.35, angle = 0]{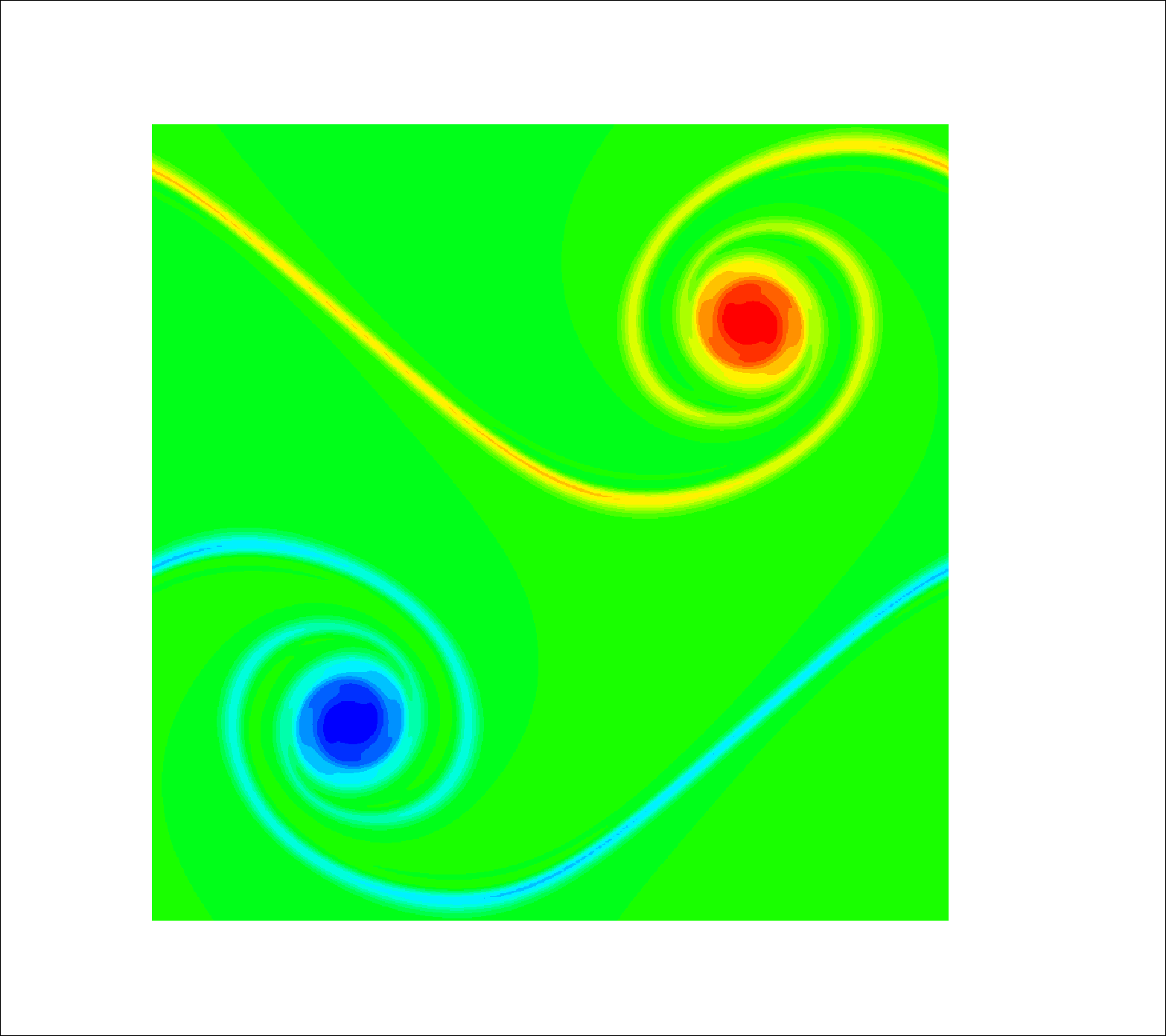}
	\caption{
		The vorticity at $t=1.0$ for the doubly periodic shear layer
		at $Re = 10000$ by the present simulation using SMAMEs (left)
		and by using the original MAMEs (right).
		The simulation parameters are
		$N_{L} = 320$, $N_{t} = 16000$ ($c = 50$).
	}
	\label{fig:dpsl-vort-t1}
\end{figure}

The third case is the doubly periodic shear layer in 2D.
The domain is a square $[0, 1] \times [0, 1]$ with periodic boundary conditions in both the $x-$ and $y-$directions.
The initial density and velocity fields are given by,
\begin{subequations}
\begin{equation}\label{eq:dpsl-rho-t0}
d(x,y,0) = 1.0 ,
\end{equation}
\begin{equation}\label{eq:dpsl-u-t0}
u(x, y, 0) = \left\{  \begin{matrix}
\tanh[\delta_{w} (y - 0.25)] && \textrm{if} \quad y \leq 0.5, \\
\tanh[\delta_{w} (0.75 - y )] && \textrm{if} \quad y > 0.5, \\
\end{matrix}
\right .
\end{equation}
\begin{equation}\label{eq:dpsl-v-t0}
v (x, y, 0) = \delta_{p} \sin (2 \pi (x + 0.25))  ,
\end{equation}
\end{subequations}
where $\delta_{w} = 80$ and $\delta_{p} = 0.05$ are two parameters related to the width of the shear layer
and the initial perturbation amplitude.
The Reynolds number is taken to be $Re= 10000$.
The simulation is performed from $t=0$ to $1$ by using the present SMAMEs
and also the original MAMEs in~\cite{jcp2020mame}.
The total enstrophy $\Omega$ and energy $E$ were calculated during the simulation as,
\begin{equation}\label{eq:dpsl-enstropy-energy}
\Omega =\frac{1}{2 S} \sum_{i,j} \omega_{i,j}^{2} dS_{i,j}, \quad
E = \frac{1}{2 S}  \sum_{i,j} (u_{i,j}^{2} +  v_{i,j}^{2} ) dS_{i,j},
\end{equation}
where $S$ is the total area of the domain, $dS_{i,j}$ is the area of the cell labelled by the indices $(i,j)$,
$\omega_{i,j}$, $u_{i,j}$, and $v_{i,j}$ are the vorticity, $x-$velocity and $y-$velocity components at the node $(i,j)$, and the summation is performed over all the nodes within the domain.
Figure \ref{fig:cmp-dpsl-enstrophy-ke-evol} shows the evolutions of the enstrophy and energy by using the SMAMEs, the MAMEs and some references results from~\cite{pre13-edac, jcp97-dpsl-test} (using the EDAC
and pseudospectral methods respectively).
Overall, the present results on a $320 \times 320$ grid are in good agreement with the pseudospectral results
obtained on a much finer ($768 \times 768$) grid.
For the enstrophy, both the results by the SMAMEs and MAMEs agree very well with the pseudospectral results (one can hardly see any differences between them from fig. \ref{fig:cmp-dpsl-enstrophy-ke-evol})
For the kinetic energy, the evolution by the MAMEs match the pseudospectral results slightly better than that by the SMAMEs.  
In addition, fig. \ref{fig:dpsl-vort-t1} shows the contours of the vorticity at $t=1$ by the present simulation
and by the MAMEs.
One can see that the two sets of results look similar to each other,
and that the curled shear layers still look smooth and there are no spurious vortices.
Due to the simplified formulation and implementation, the present simulation only takes about $258.6$s
whereas that using the original MAMEs takes $365.5$s on the same computer under the same settings.
That means the present method saves about $30\%$ computation time compared with the original method using the MAMEs.
It is noted that the present simulation is unstable when the mesh is too coarse. 
Even on a $256 \times 256$  grid the simulation blowed up 
(the simulation using the MAMEs was not stable, either).
Previously, it was found that some more robust upwind methods can keep the simulation at such a high $Re$ stable on a coarse grid. However, they often produce spurious vorticities under such situations~\cite{jcp97-dpsl-test}. 
The present method using the SMAMEs has low tolerance to the under-resolved situations.
On the other hand, it is less likely to produce spurious vortices and unphysical results.

\subsection{Capillary wave in 2D}\label{ssec:cap-wave}

Next, we study some more complicated two-phase flows using the CHE for interface capturing. 
Besides, the surface tension effects are taken into account in the momentum equations.
For such problems, there are additional parameters: (1) the Cahn number $Cn = W/L_{c}$ (i.e., the interface thickness measured by the characteristic length) and (2) the Peclet number $Pe = U_{c} L_{c}^{2} / (M \sigma)$ (reflecting the relative importance of convection over diffusion in the CHE).
For two-phase flows one can derive a velocity scale from the surface tension and viscosity as $\sigma / (\rho_{0} \nu)$ and it is chosen to be the default characteristic velocity $U_{c}$.
From $U_{c}$ one can derive a characteristic time as $T_{c} = L_{c} /U_{c}$.
For two-phase problems, the above $U_{c}$ and $T_{c}$ are used to scale the velocity and time (unless specified otherwise).

The first two-phase problem is the 2D capillary wave.
The domain is a square $[0, 1] \times [0, 1]$ ($L_{c}$ is set to the side length $L = L_{x} = L_{y}$).
The left and right boundaries are periodic, and the top and bottom boundaries are no-slip walls.
The upper half domain is filled with the "red" fluid where $\phi = 1$ and 
the lower is filled with the "blue" fluid where $\phi = -1$ 
(note that the two fluids have the same density and viscosity, thus are completely symmetric; 
for convenience we denote them as "red" and "blue").
The initial interface is slightly perturbed with the interface position varying with $x$ as
$h (x) = h_{eq} + A_{p} \cos[ k (x + 0.5)]$,
where $h_{eq} = 0.5$ is the equilibrium interface position,
$A_{p} = 0.01$ is the amplitude of disturbance and 
$k = 2 \pi / \lambda$ is the wavenumber
($\lambda = L_{x} = 1$ is the wavelength).
The initial order parameter field is set to
$\phi(x,y, 0) = \tanh [ 2 (y - h(x) )  / Cn]$.
The interface position $h_{0}(t)$ at $x=0$ was monitored during the simulation.
The Reynolds number is defined as $Re = U_{c} L_{c} / \nu = \sigma L_{c}/ (\rho_{0} \nu^{2})$.
The case at $Re = 1000$ was first investigated.
For this problem, there exists a basic frequency $\omega_{0} = \sqrt{\sigma k^{3} / (2 \rho_{0} )}$.
When both the liquid and gas have the same the kinematic viscosity ($\nu$) and the perturbation is small ($A_{p} \ll 1$),
one can obtain the analytical solution for this problem~\cite{phf81:gravcapwav, kim05:cstff},
\begin{equation}\label{eq:cap-wave-analytical}
\tilde{h} (t) = \frac{h_{eq} - h_{0} (t)}{A_{p}} 
= \sum_{i = 1}^{4} \frac{z_{i} \omega_{0}^{2}}{Z_{i} (z_{i}^{2} - \bar{\epsilon} \omega_{0}) } 
\exp \bigg[\frac{(z_{i}^{2} - \bar{\epsilon} \omega_{0}) t'} {\omega_{0}}\bigg]  \text{erfc}\bigg(z_{i} \sqrt{\frac{t'}{\omega_{0}}}\bigg) ,
\end{equation}
where $t' = \omega_{0} t $ and $\bar{\epsilon} = \nu k^{2} / \omega_{0}$
are the scaled time and dimensionless viscosity, 
$z_{i}$ are the four roots of the algebraic equation
$z^{4} -  \sqrt{\bar{\epsilon} \omega_{0}} z^{3} - \bar{\epsilon} \omega_{0} z^{2} +  (\bar{\epsilon} \omega_{0})^{\frac{3}{2}} z + \omega_{0}^{2} = 0$
and $Z_{1} = (z_{2} - z_{1}) (z_{3} - z_{1}) (z_{4} - z_{1})$,
$Z_{2} = (z_{3} - z_{2}) (z_{4} - z_{2})(z_{1} - z_{2}) $,
$Z_{3} = (z_{4} - z_{3}) (z_{1} - z_{3}) (z_{2} - z_{3})$,
$Z_{4} = (z_{1} - z_{4}) (z_{2} - z_{4})(z_{3} - z_{4}) $.
Figure \ref{fig:cmp-dev-u-evol}a shows the evolutions of $\tilde{h} $  
over $0 \leq t \leq 40$ by the present method and the MRT-LBM
using the same numerical parameters $N_{L} = 64$, $N_{t} = 384$ ($c = 6$).
It is seen that the present numerical results are very close to (almost overlap) that by the MRT-LBM. Both numerical solutions agree with the analytical one in the early stage and the deviations increase gradually with time.
After about two oscillation periods, the deviations remain to be small and can actually be reduced by increasing the resolution in space and time.
This is observed from fig. \ref{fig:cmp-dev-u-evol}b which also shows the results obtained by the present method using a finer mesh with $N_{L} = 128$, $N_{t} = 3072$ ($c = 24$).
Besides, another case at an even higher $Re$($= 4000$) was studied.
Table \ref{tab:cmp-period-cap-wave} compares the oscillation periods obtained by the present simulations
using two sets of meshes with the analytical periods for the two cases.
It is seen that under all situations the deviations in the period are small (less than $5 \%$), 
and as the grid is refined ($N_{L}$ is changed from $64$ to $128$) the deviations decrease quickly to around $ 1 \%$.
Finally, it is noted that different values of $N_{t}$($c$) were tested for the case at $Re = 1000$ with
$Cn = 0.0625$ and $Pe = 2 \times 10^{4}$.
When $c$ further increases (to $10$ and $20$), the results are almost the same as that obtained with $c=6$.
When $c$ decreases to $5$, the simulation becomes unstable no matter whether the present SMAMEs or the LBM is used. The reason is likely to be that for two-phase flows the CHE for interface evolution may impose an even more stringent condition on the time step. Under such situations, the present method should be as robust as the LBM with regard to the stability issue. At the same time, it is much easier to implement and performs faster. Therefore, overall the present method can be more competitive than the LBM for two-phase flows.

\begin{figure}[htp]
	\centering
	\includegraphics[trim= 1mm 1mm 1mm 1mm, clip, scale = 0.65, angle = 0]{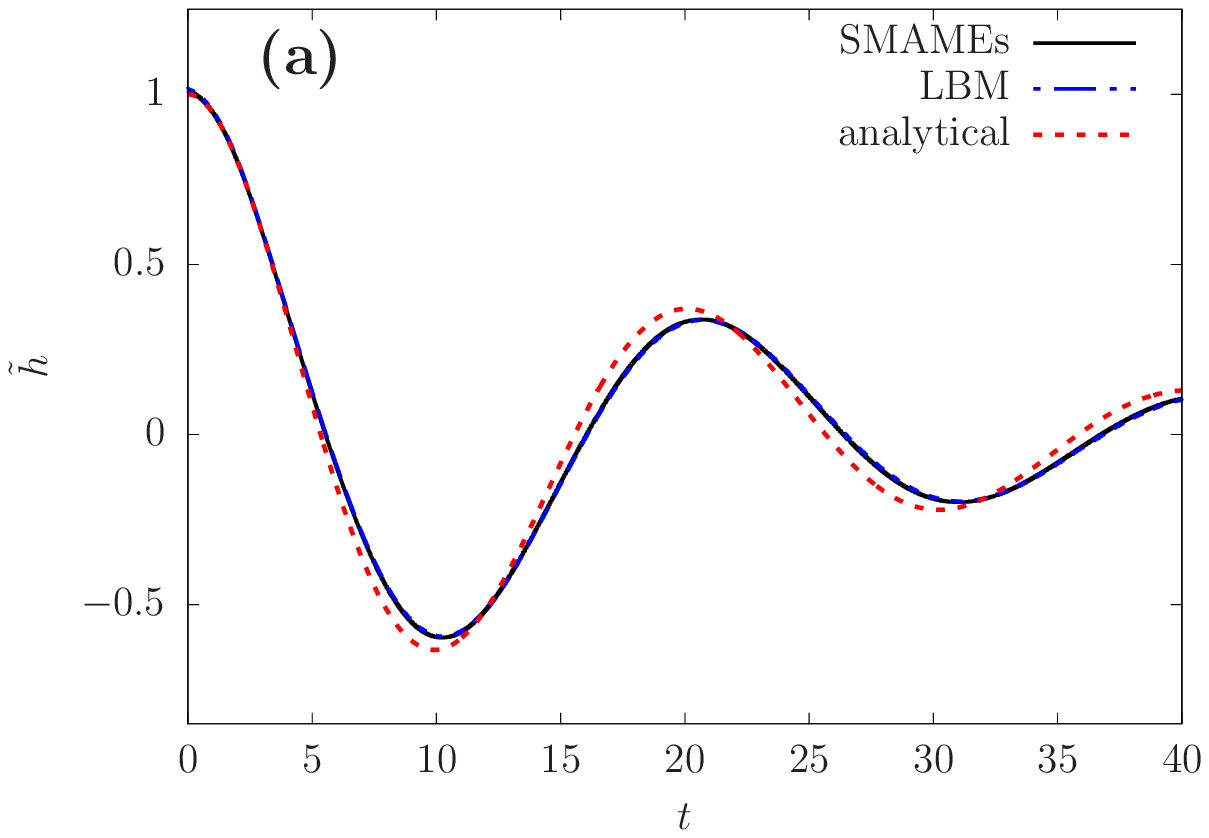}
	\includegraphics[trim= 1mm 1mm 1mm 1mm, clip, scale = 0.65, angle = 0]{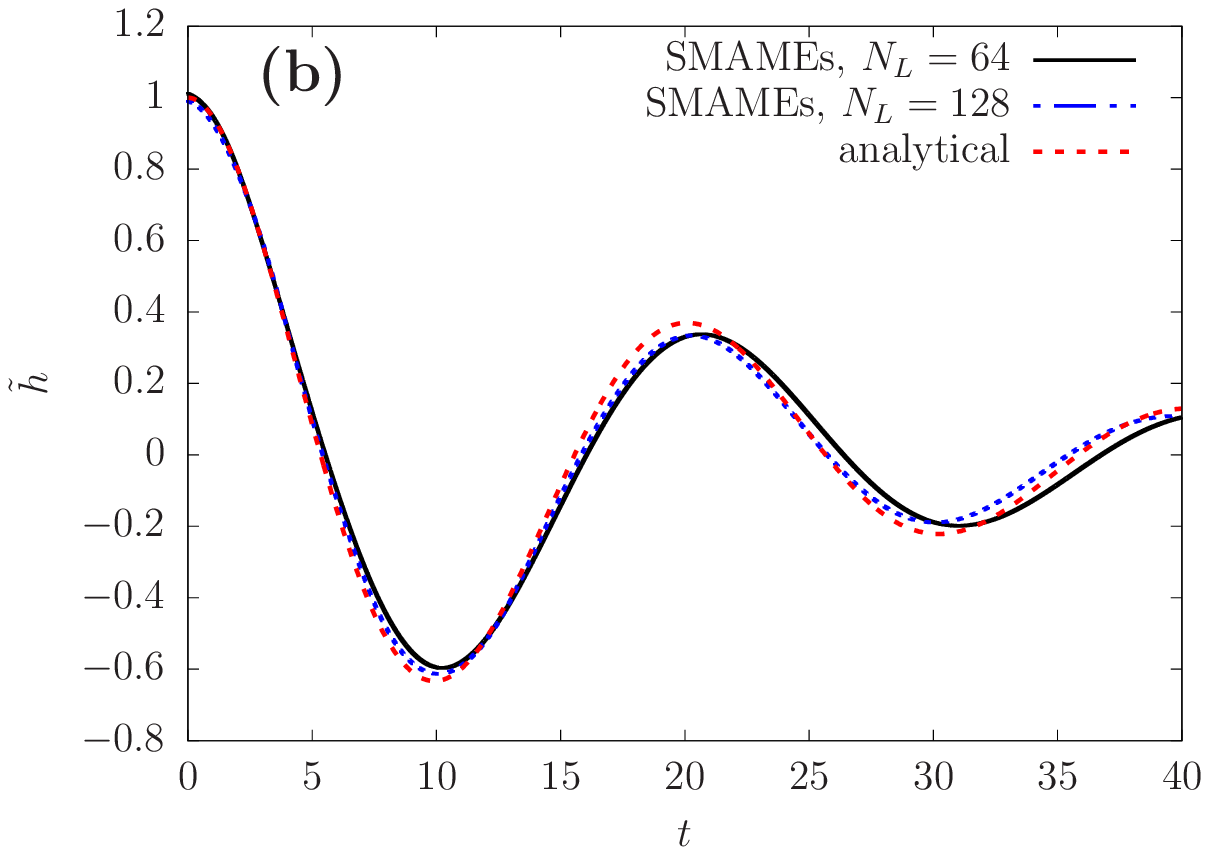}
	\caption{
		Evolutions of the (scaled) interface displacement at the left boundary for the capillary wave  
		 at $Re = 1000$.
		In both (a) and (b), the dashed line is the analytical prediction by eq. \ref{eq:cap-wave-analytical}.
		In (a), the solid line is by the present method, and the dash-dotted line is by the MRT-LBM.
		In (b), the solid line is obtained at $N_{L} = 64$, and the dash-dotted line, $N_{L} = 128$.
		The shared simulation parameters in (a) are
		$N_{L} = 64$, $N_{t} = 384$ ($c = 6$), $Cn = 0.0625$, $Pe = 2 \times 10^{4}$.
		For the fine mesh solution in (b), $N_{t} = 3072$ ($c = 24$), $Cn = 0.03125$, $Pe = 2 \times 10^{4}$.
	}
	\label{fig:cmp-cap-wave-evol}
\end{figure}

\begin{table}
	\begin{center} 
		\begin{tabular}{|c|c|c|} \hline
			Reynolds number & 1000 & 4000 	\\\hline
			Period (analytical) &  20.071 & 38.751 	\\\hline
	        Period $\vert$ Error ($N_{L} = 64$, $c=6$) &  20.669 $\vert$ 2.98\% & 40.089 $\vert$ 3.45\%	\\\hline
	        Period $\vert$ Error ($N_{L} = 128$, $c=24$) &  20.265 $\vert$ 0.97\% & 39.293  $\vert$ 1.40\%	\\\hline
		\end{tabular}
	\end{center}
	\caption{Comparisons of the oscillation period for the capillary wave
		at $Re = 1000$ and $4000$.
	The shared simulation parameter is $Pe = 2\times 10^{4}$. The Cahn numbers are $Cn = 0.0625$ for $N_{L} = 64$
and $Cn = 0.03125$ for $N_{L} = 128$.}
	\label{tab:cmp-period-cap-wave}
\end{table}

\subsection{Falling drop}\label{ssec:falling-drop}

\begin{figure}[htp]
	\centering
	\includegraphics[trim= 1mm 0mm 1mm 1mm, clip, scale = 0.65, angle = 0]{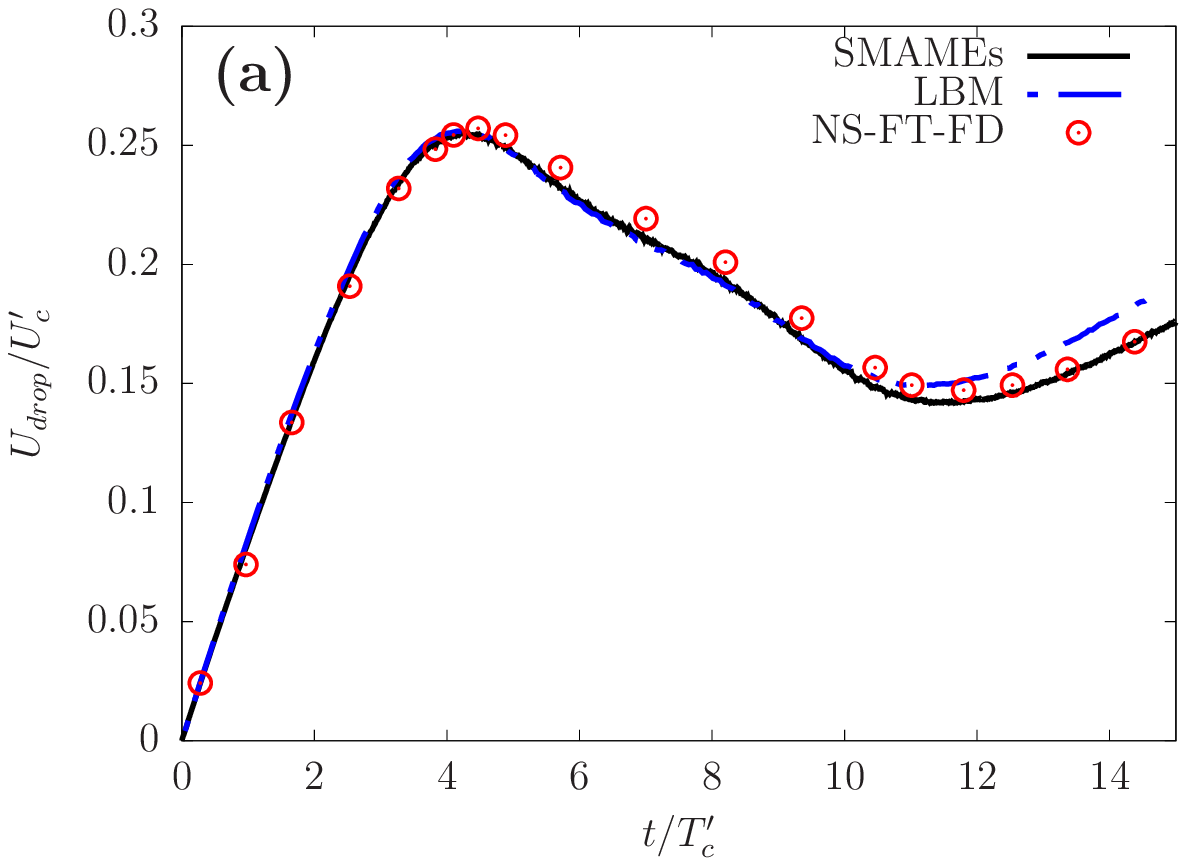}
	\includegraphics[trim= 1mm 0mm 1mm 1mm, clip, scale = 0.65, angle = 0]{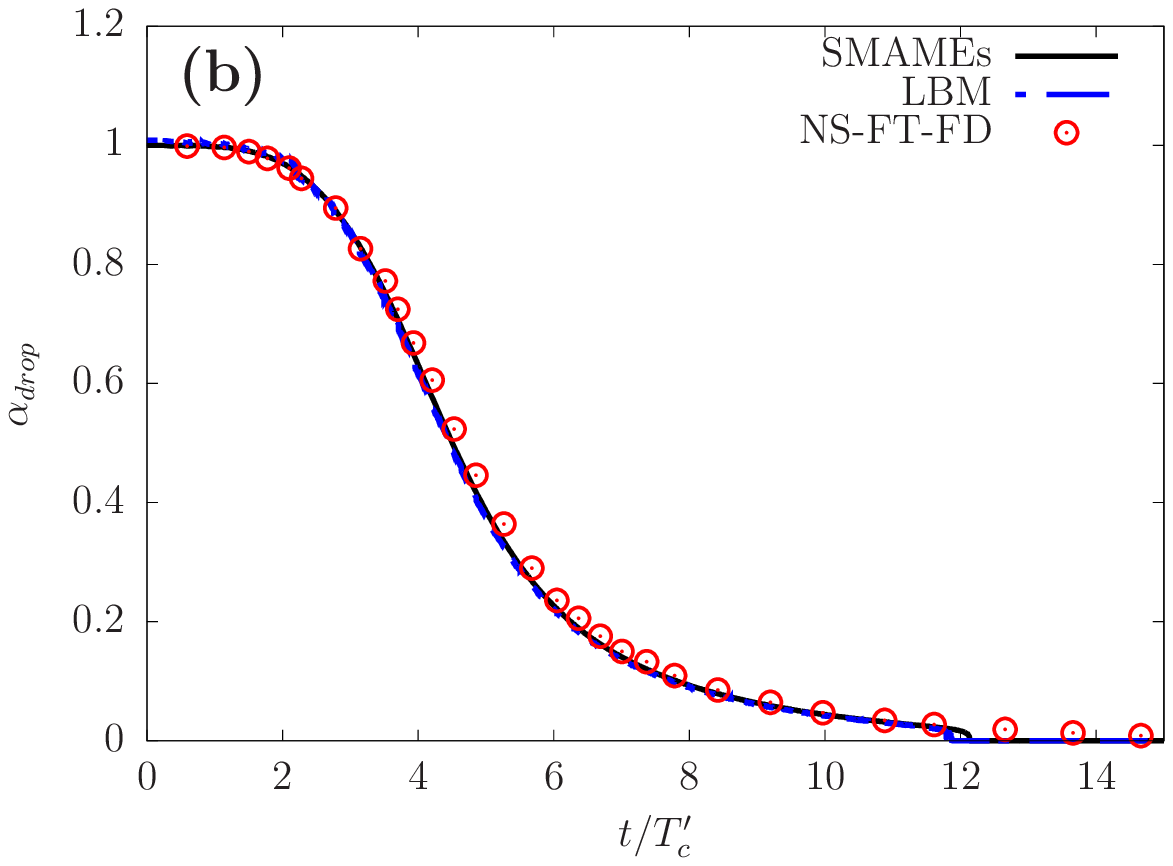}
	\caption{
		Evolutions of (a) the centroid velocity $U_{drop}$ in the axial direction and (b) the aspect ratio $\alpha_{drop}$ of the drop at  $Eo = 144$ and $Oh = 0.0466$ for the falling drop problem
		by the present method, by the axisymmetric LBM in~\cite{hybrid-mrt-lb-fd-axisym} and by the finite difference solution of the NSEs and the front-tracking method~\cite{pof99dropacc}.
		The present simulation parameters are
		$N_{L} = 50$, $N_{t} = 4000$ ($c = 80$), $Cn=0.06$ and $Pe = 1000$.
	}
	\label{fig:falling-drop-evol}
\end{figure}

The second two-phase problem is a falling drop under the action of a body force.
This problem is symmetric about the $z$-axis and can be simplified to an axisymmetric problem.
Previously, it was studied by an axisymmetric LBM in~\cite{hybrid-mrt-lb-fd-axisym} and by a finite difference front tracking method in~\cite{pof99dropacc} that solves the incompressible NSEs.
In this problem, a drop is surrounded by the ambient gas. 
The drop/gas density ratio is $r_{\rho} = \rho_{L}/\rho_{G} = 1.15$ 
($\rho_{L}$ and $\rho_{G}$ are the densities of the liquid and gas)
and the dynamic viscosity ratio is $r_{\eta} = \eta_{L} / \eta_{G} = 1$
($\eta_{L}$ and $\eta_{G}$ are the dynamic viscosities of the liquid and gas).
The drop radius $R$ is chosen as the characteristic length ($L_{c} = R$).
The domain is a rectangle $[0, 24] \times [0, 8]$ ($L_{z} = 24$ and $L_{r} = 8$).
Symmetric boundary conditions are applied on the boundary $r = 0$ and no slip wall boundary conditions are used for the other three boundaries.
The initial drop center is located at $(z_{c}, r_{c}) = (2, 0)$.
The order parameter field is initialized to be $\phi(z,r, 0) = - \tanh [ 2 (r_{dc} - R )  / Cn]$ where $r_{dc}= \sqrt{(z - z_{c})^{2} + (r - r_{c})^{2}}$.
The body force of magnitude $g$ is applied along the $z-$direction.
Note that with some manipulation of the pressure, the body force may be applied only on the drop~\cite{cmfd99book}.
Two main dimensionless parameters are the Eotvos number and Ohnesorge number defined as,
\begin{equation}\label{eq:falling-drop-dimless-num}
Eo = \frac{g (\rho_{L} - \rho_{G}) D^{2}}{\sigma}, \quad Oh = \frac{\eta_{L}}{\sqrt{\rho_{L} D \sigma}} ,
\end{equation}
where $D = 2R $ is the drop diameter.
They are set to $Eo = 144$ and $Oh = 0.0466$ (same as in~\cite{hybrid-mrt-lb-fd-axisym, pof99dropacc}).
To facilitate the comparison with previous results, we scale the velocity and time using
$U_{c}^{\prime} = \sqrt{g D}$ and $T_{c}^{\prime} = \sqrt{D/g}$.
Because the density ratio is small, the Boussinesq approximation is used here (as in~\cite{hybrid-mrt-lb-fd-axisym}),
and the physical density is assumed to be unity for both fluids.
To account for the density difference, one needs to multiply the body force acting on the drop by a factor $(\rho_{L} - \rho_{G}) / [(\rho_{L} + \rho_{G}) / 2] = 2 (r_{\rho} - 1) / (r_{\rho} + 1)$ where $(\rho_{L} + \rho_{G}) / 2$ is the (real) average density of the two fluids (see the Appendix of~\cite{hybrid-mrt-lb-fd-axisym}).
During the simulation, we monitor the centroid velocity $U_{drop}$ along the axial direction,
the drop thickness (in the axial direction) $Th_{drop} $ and the drop width (in the radial direction)  $ Wh_{drop}$. 
From the latter two, we calculate the aspect ratio of the drop as $\alpha_{drop} = Th_{drop} / Wh_{drop}$. 
The centroid velocity $U_{drop}$ is calculated 
by $U_{drop} = \int_{A \vert_{\phi > 0}} r u(r,z)dr dz / \int_{A \vert_{\phi > 0}} r dr dz $ 
where $A \vert_{\phi > 0}$ represents the region where $\phi > 0$.
Figure \ref{fig:falling-drop-evol}
shows the evolutions of the centroid velocity $U_{drop}$ and the aspect ratio $\alpha_{drop}$ of the drop obtained by the present method together with those from~\cite{hybrid-mrt-lb-fd-axisym, pof99dropacc}.
It is found that the present result follows the prediction by the axisymmetric LBM very well,
and both of them are close to that by the front tracking method in~\cite{pof99dropacc} till
$t/T_{c}^{\prime} \approx 12$.
After that, the front tracking method still predicts a non-zero drop thickness whereas the simulations using the phase field model (both the present and ~\cite{hybrid-mrt-lb-fd-axisym}) predict drop breakup. This is an inherent difference between the two types of methods.

\subsection{Drop spreading and dewetting on a wall}\label{ssec:drop-dewetting}

\begin{figure}[htp]
	\centering
	\includegraphics[trim= 1mm 1mm 1mm 1mm, clip, scale = 0.65, angle = 0]{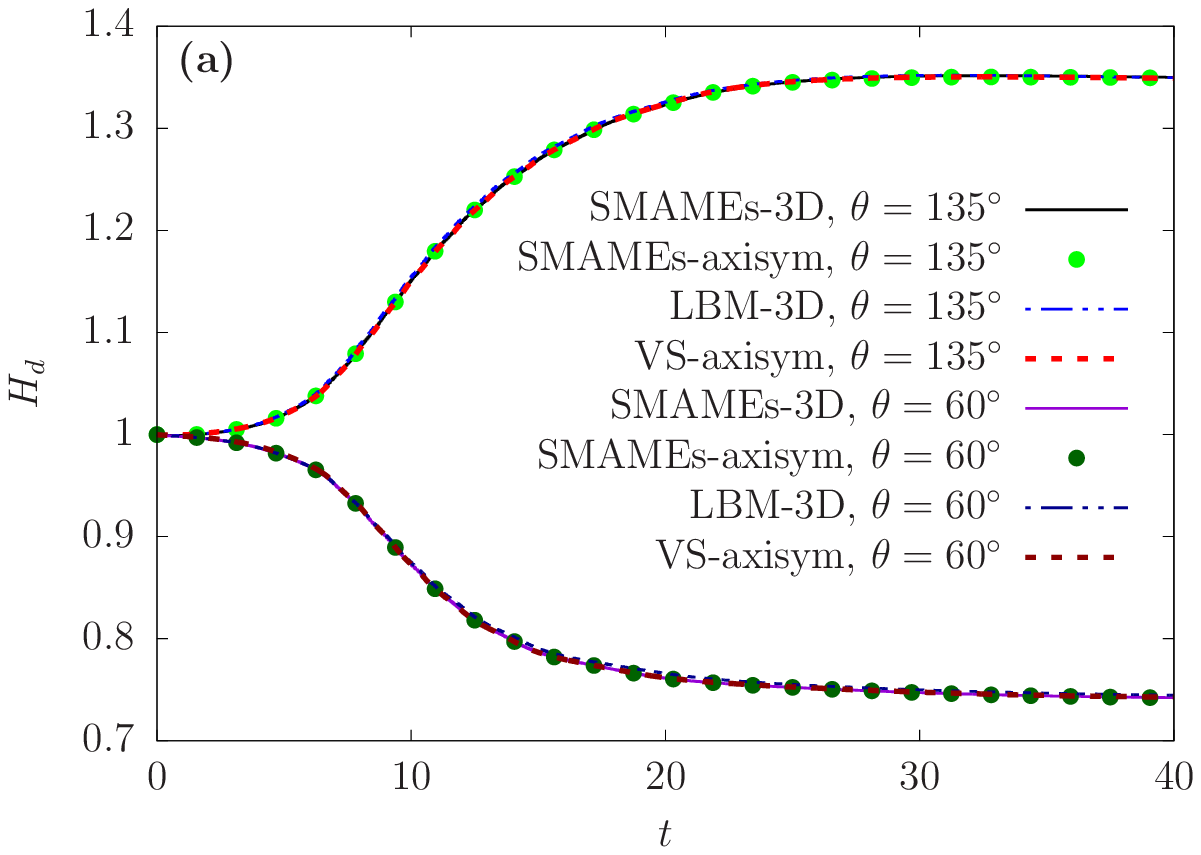}
	\includegraphics[trim= 1mm 1mm 1mm 1mm, clip, scale = 0.65, angle = 0]{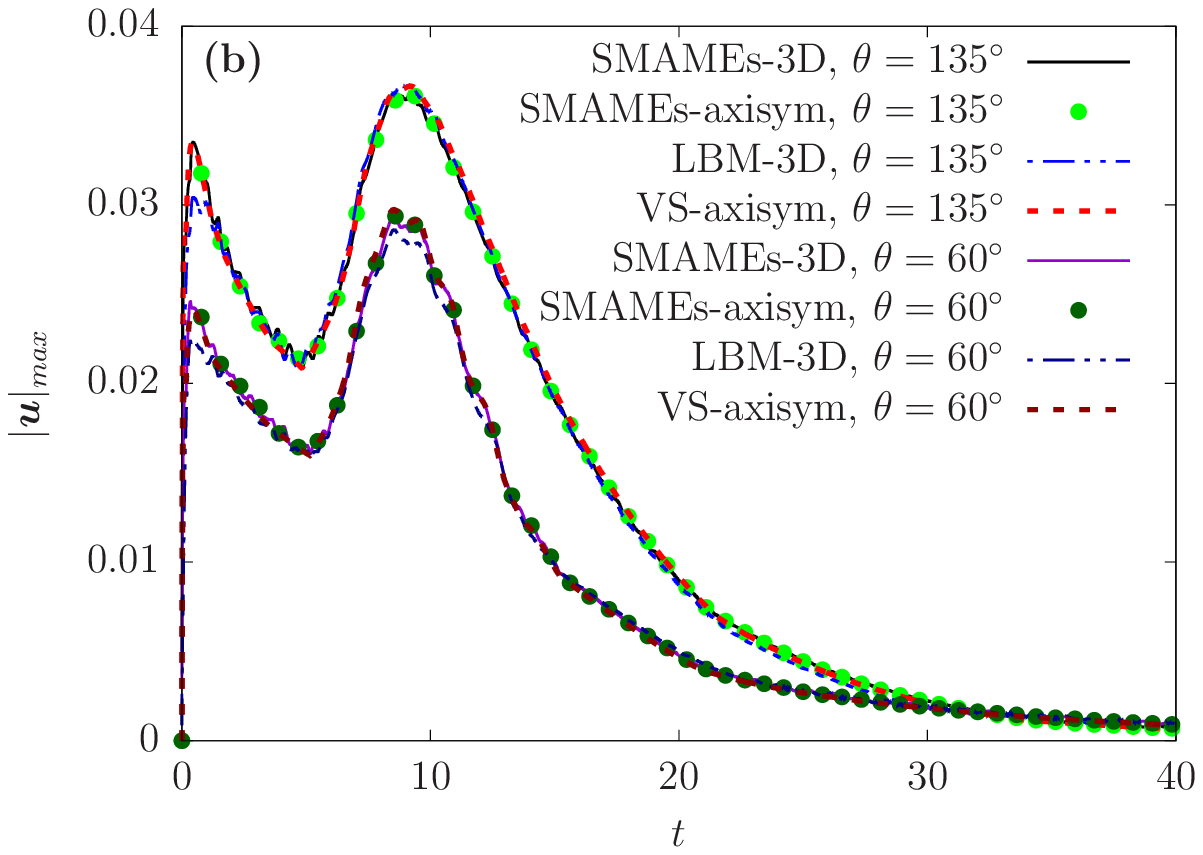}
	\caption{
		Evolutions of (a) the droplet height $H_{d}$ 
		(b) the maximum velocity magnitude $\vert \boldsymbol{u} \vert _{max}$ during the drop dewetting / spreading computed by three different methods.
		The solid lines are by the present 3D simulations, the filled circles are by the present axisymmetric simulations, the dash-dot-dot lines are by the 3D LBM using the weighted MRT collision model~\cite{jcp17-wmrt-lbm-mfflow}
		and the dashed lines are by the axisymmetric simulation using the VS formulation~\cite{pcfd15-vs-pf}.  
		The shared simulation parameters are
		$N_{L} = 32$, $N_{t} = 320$ ($c = 10$), $Cn = 0.1$ and $Pe = 8000$.
	}
	\label{fig:cmp-drop-height-vmmax}
\end{figure}

The third two-phase problem is on the motion of a drop on a wall with a contact angle $\theta$. 
Initially, the drop is a hemisphere with a radius $R = 1$ (i.e., the characteristic length $L_{c} = R$) 
and its center at $(0,0,0)$.
The Ohnesorge number is given by $Oh  = \rho_{0} \nu / \sqrt{\rho_{0} R \sigma} = 1 /\sqrt{Re}$. 
Two cases with $\theta = 135^{\circ}$ and $60^{\circ}$ at $Re = 100$ ($Oh = 0.1$) were investigated.
On the hydrophobic wall with $\theta = 135^{\circ}$ the drop dewets from the wall
whereas on the hydrophilic wall with $\theta = 60^{\circ}$ the drop spreads on the wall.
The domain size is a cube $[-3,3] \times [-3,3] \times [0,3]$.
Due to symmetry, the actual simulation domain was $[0,3] \times [0,3] \times [0,3]$.
Symmetric boundary conditions were applied on four side boundaries ($x = 0, 3$ and $y = 0, 3$),
and no slip boundary conditions were used on the top and bottom boundaries ($z=0, 3$).
This problem is actually symmetric about the $z-$axis and 
may be also handled under the axisymmetric geometry.
In the axisymmetric simulations using the SMAMEs, 
the domain is $[0,3] \times [0,3]$ ($L_{z} = 3$ and $L_{r} = 3$).
For the LBM simulations, the D3Q19 velocity model is used and the collision model is the weighted MRT~\cite{jcp17-wmrt-lbm-mfflow}.
Besides, axisymmetric simulations using the vorticity stream function (VS) formulation~\cite{pcfd15-vs-pf}
were also performed. 
Unlike the artificial compressibility methods (e.g., LBM, MAMEs or SMAMEs),
the VS formulation solves the incompressible NSEs without any compressibility errors
(it has to solve the Poisson-like equations). 
Two main quantities were monitored: the drop height (in the $z-$direction) $H_{d}$ and the maximum velocity magnitude $\vert \boldsymbol{u} \vert _{max}$ in the whole domain.
Figure \ref{fig:cmp-drop-height-vmmax} shows the results by the four sets of simulations
(the 3D and axisymmetric SMAMEs, 3D LBM and axisymmetric VS) for this problem
using the same $\delta_{x}$ and $\delta_{t}$.
It is seen that for both cases at $\theta = 135^{\circ}$ and $60^{\circ}$ all the methods predict the evolutions of $H_{d}$ and $\vert \boldsymbol{u} \vert _{max}$ to be quite close to each other.
More careful examinations reveal that the present results by the SMAMEs are closer to those by the VS-based solver.
This could be attributed to the simplifications made in the present method which probably reduce the overall magnitude of the error terms.

\subsection{Coalescence induced drop jumping on a nonwetting wall in 3D}\label{ssec:drop-jumping}

The last problem is on the coalescence induced drop jumping on a nonwetting wall (contact angle $\theta = 180^{\circ}$) in 3D.
The domain is a box $[-3,3] \times [-3,3] \times [0,5]$.
Initially there are two spherical drops having the same radius $R$ 
(chosen as the characteristic length $L_{c}$)
and their centers are at $(\pm 1, 0, 1)$.
They start to coalesce with each other and interact with the nonwetting wall at the same time.
Unde certain conditions, the drop after coalescence may jump away from the wall~\cite{prl09jumpingdrop}.
Due to symmetry, the actual simulation domain was $[0,3] \times [0,3] \times [0,5]$
($L_{x} = L_{y} = 3$ and $L_{z}  =5$).
Because the two fluids have the same density and viscosity, 
they are better viewed as two liquid phases.
Under such conditions, the coalesced drop experiences larger drag forces and gains less momentum to jump
than water drops in air (as in the experiments~\cite{prl09jumpingdrop}).
On the other hand, it was reported that coalescence induced drop jumping can also occur when the ambient fluid is another liquid~\cite{prl15-drop-jump-fiber} (if it is on a hydrophobic fiber and the viscosity is moderate).
Here our main purpose is not to investigate the physical problem in detail.
We only intend to simulate typical cases of this interesting problem by using the proposed new method to evaluate its accuracy and efficiency.

\begin{figure}[htp]
	\centering
	\includegraphics[trim= 1mm 1mm 1mm 1mm, clip, scale = 0.8, angle = 0]{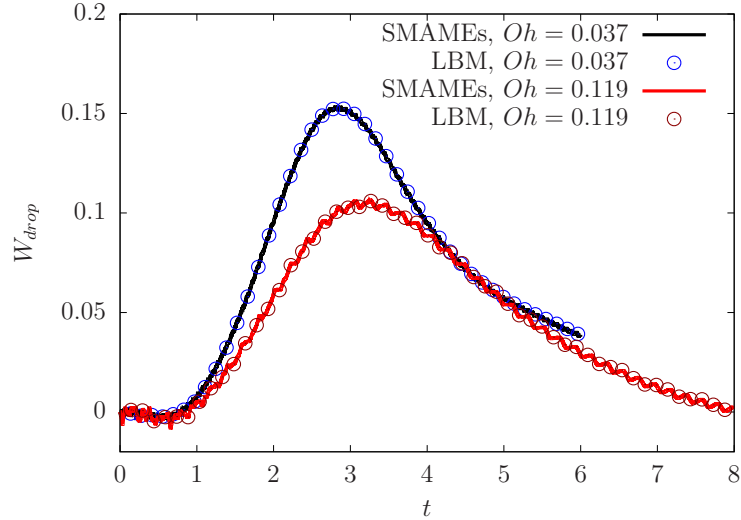}
	\caption{
		Evolutions of the $z-$component of the centroid velocity of the drop at $Oh = 0.037$ and $0.119$.
		The shared simulation parameters are
		$N_{L} = 40$, $N_{t} = 400$ ($c = 10$), $Cn = 0.1$, $Pe = 8 \times 10^{3}$.
	}
	\label{fig:jump-drop-wdrop-evol}
\end{figure}

\begin{figure}[htp]
	\centering
\includegraphics[trim= 0mm 0mm 0mm 0mm, clip, scale = 0.23, angle = 0]{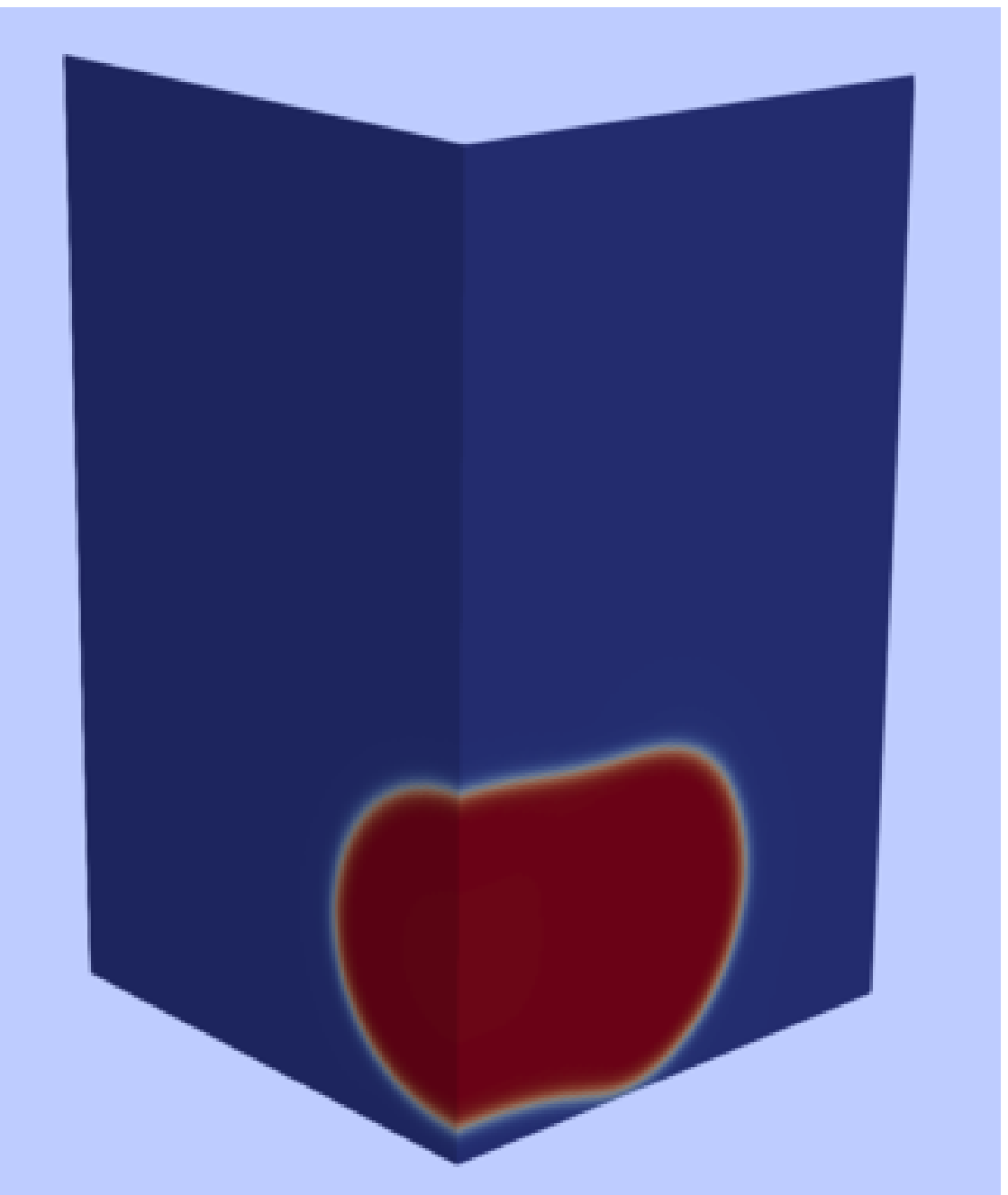}
\includegraphics[trim= 0mm 0mm 0mm 0mm, clip, scale = 0.23, angle = 0]{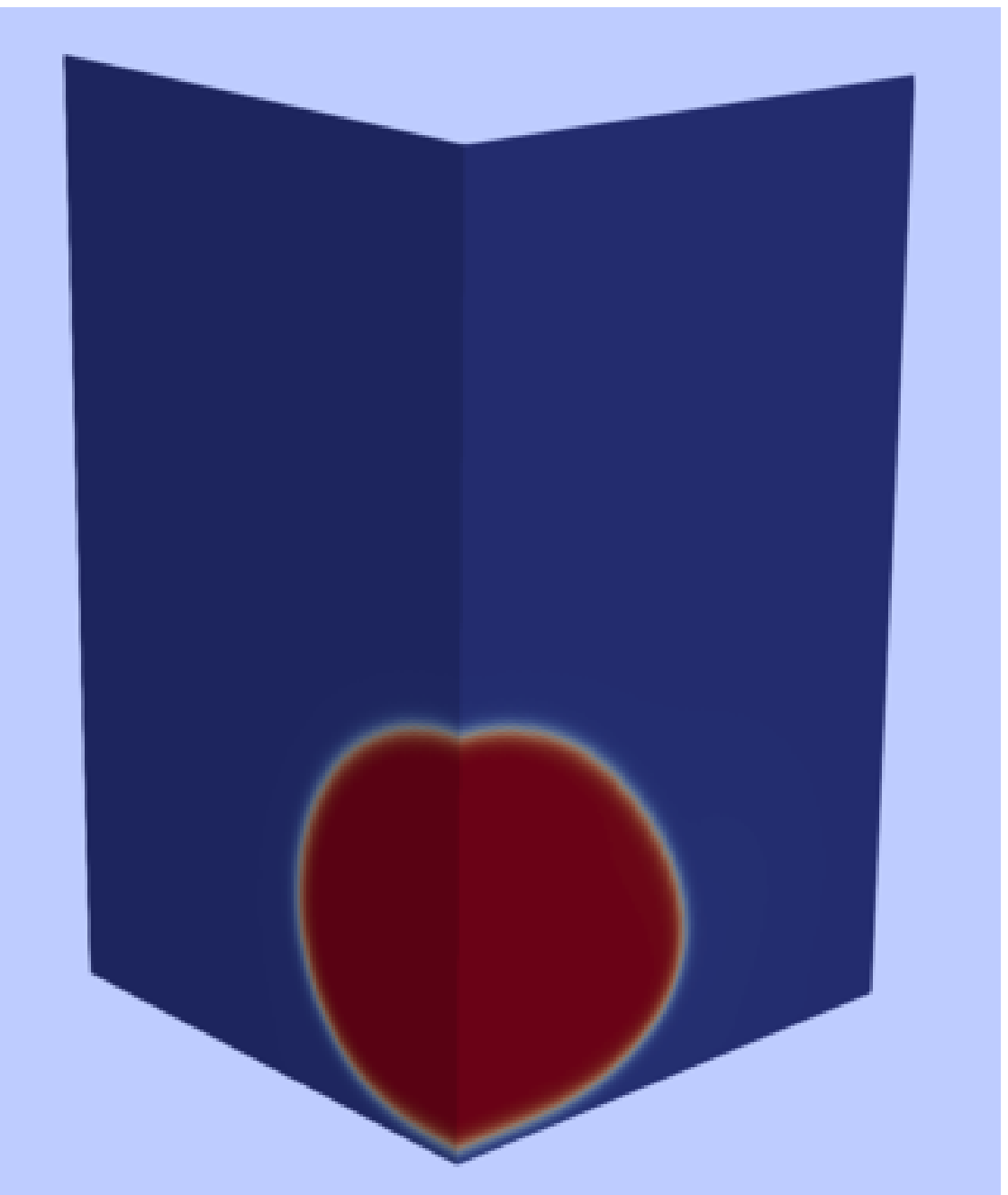}
\includegraphics[trim= 0mm 0mm 0mm 0mm, clip, scale = 0.23, angle = 0]{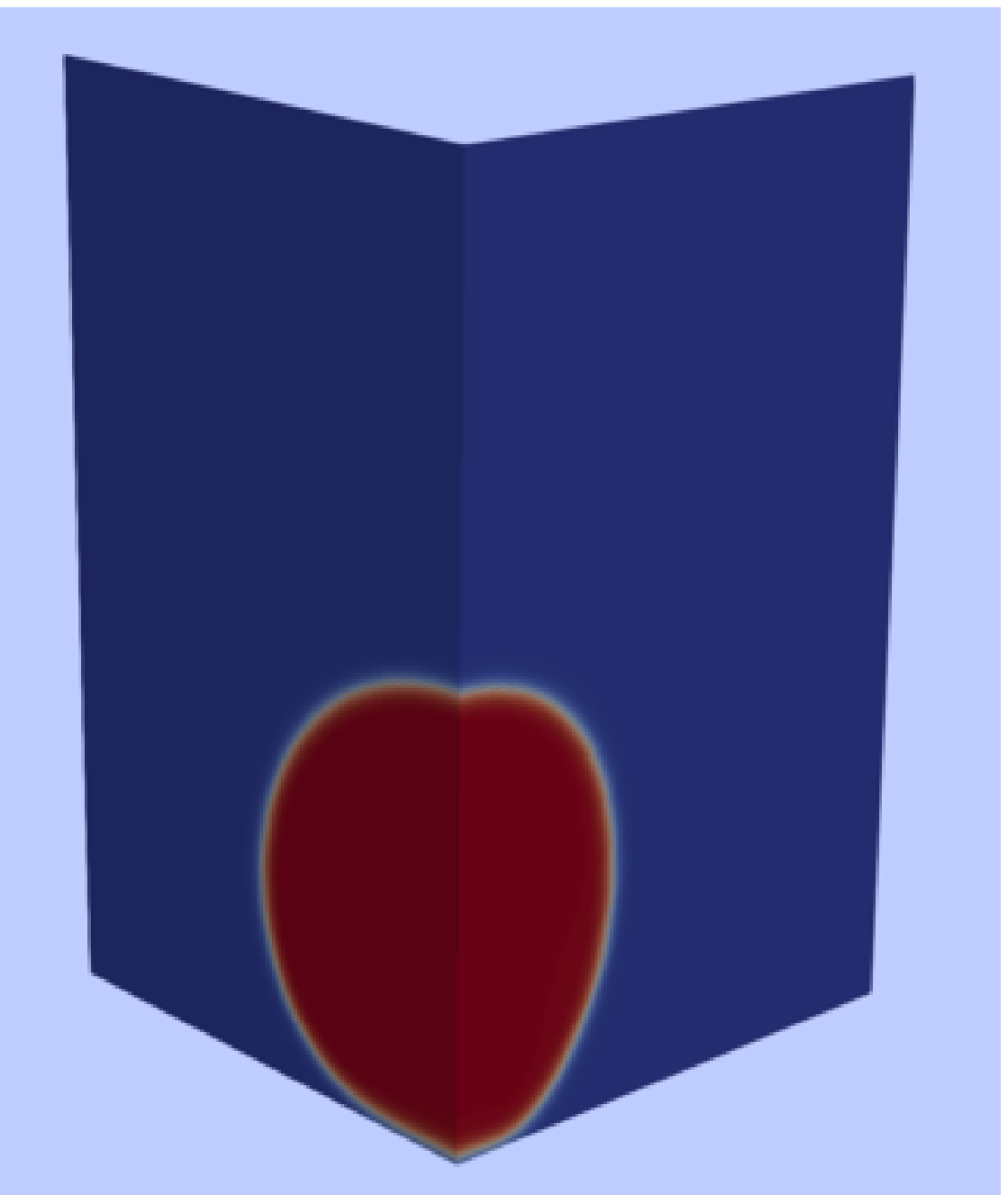}
\includegraphics[trim= 0mm 0mm 0mm 0mm, clip, scale = 0.23, angle = 0]{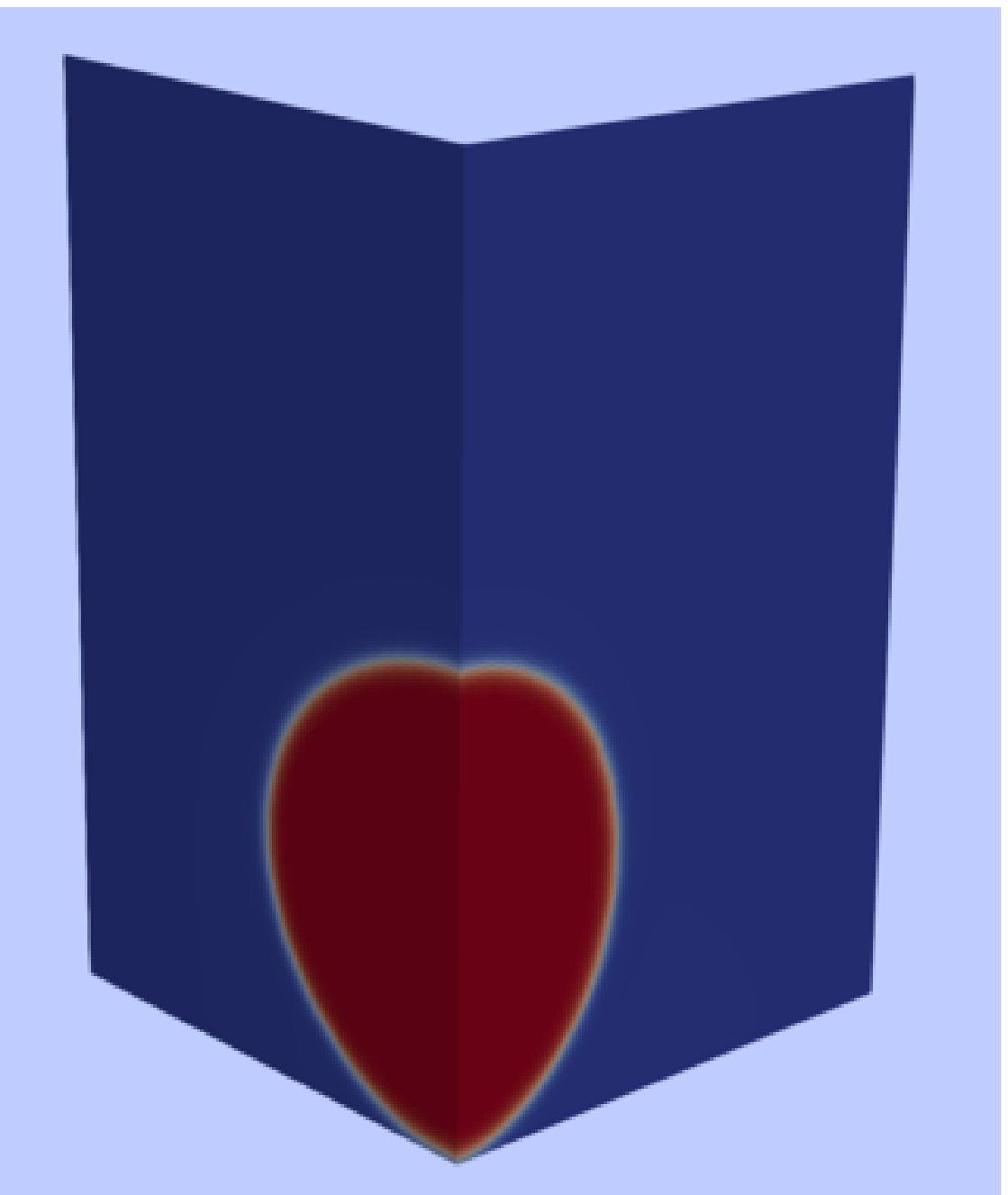}
\includegraphics[trim= 0mm 0mm 0mm 0mm, clip, scale = 0.23, angle = 0]{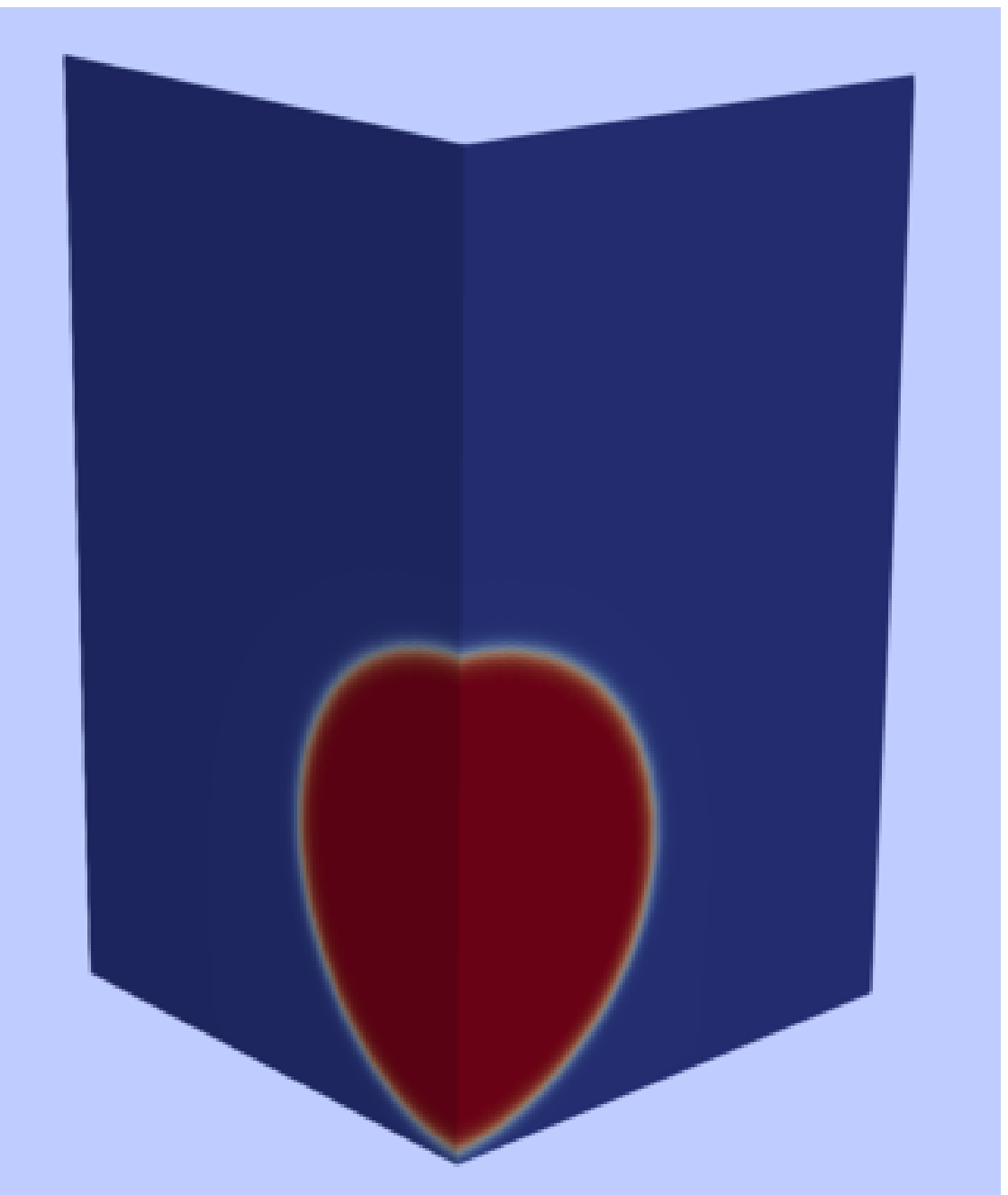}
\includegraphics[trim= 0mm 0mm 0mm 0mm, clip, scale = 0.23, angle = 0]{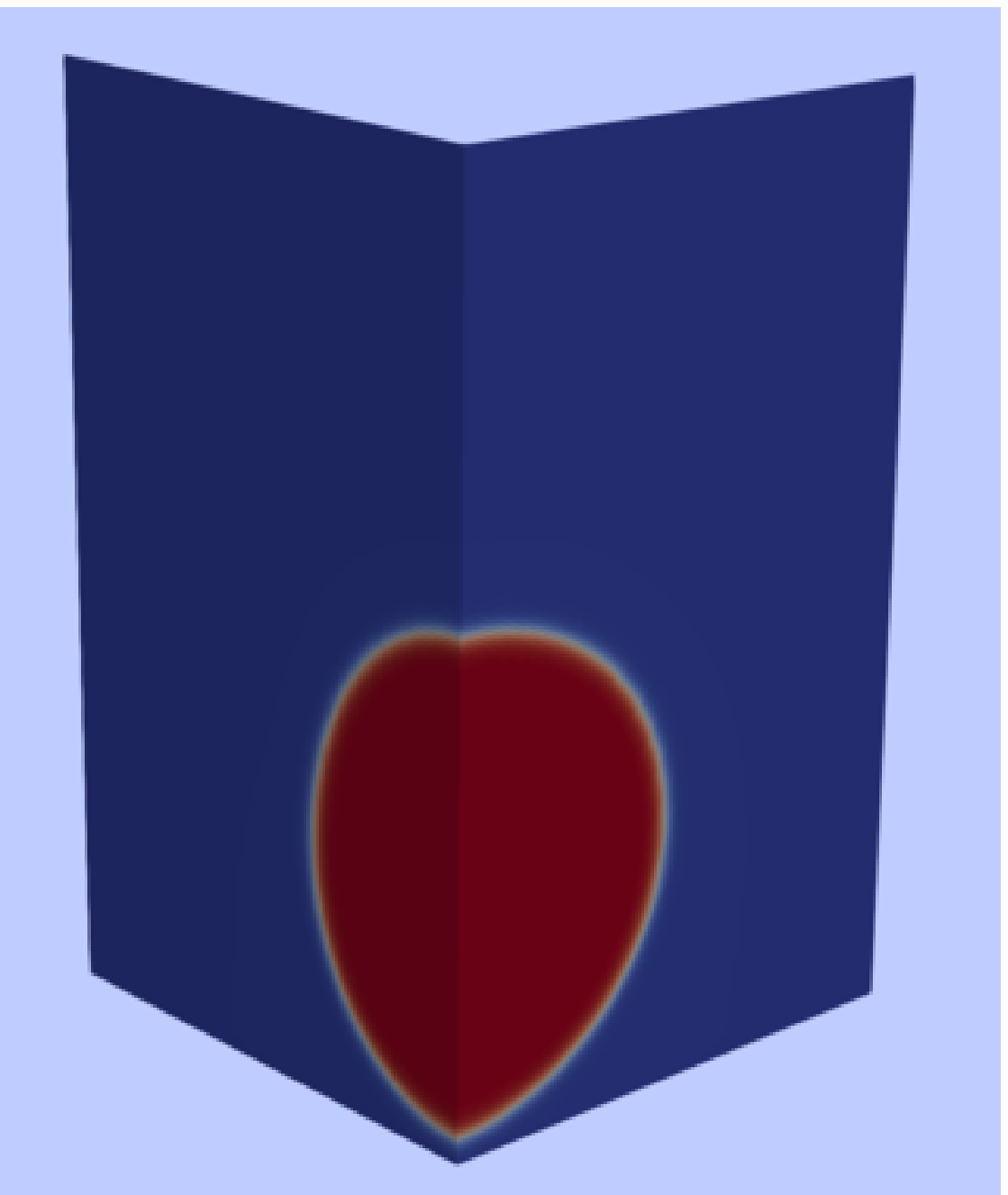}
	\includegraphics[trim= 20mm 1mm 20mm 1mm, clip, scale = 0.3, angle = 0]{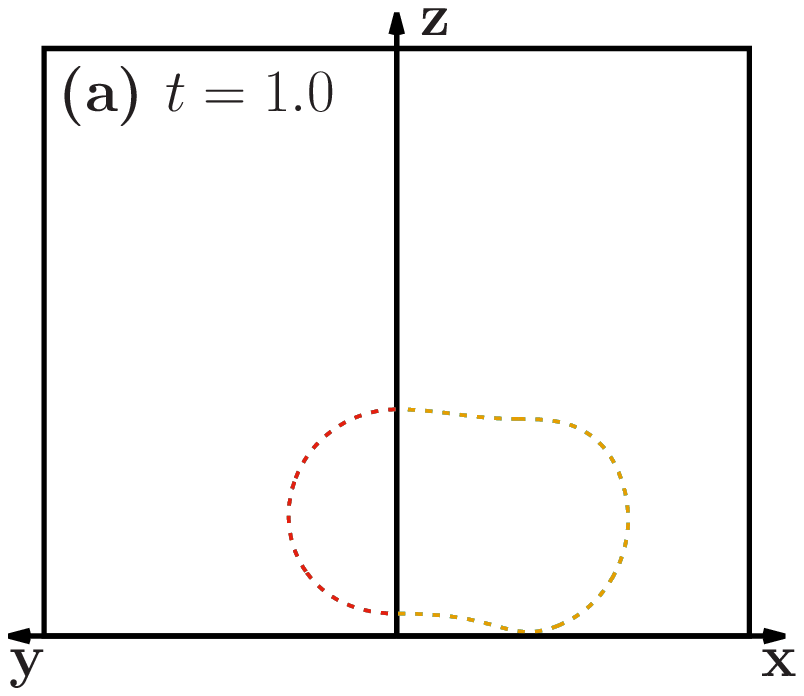}
	\includegraphics[trim= 20mm 1mm 20mm 1mm, clip, scale = 0.3, angle = 0]{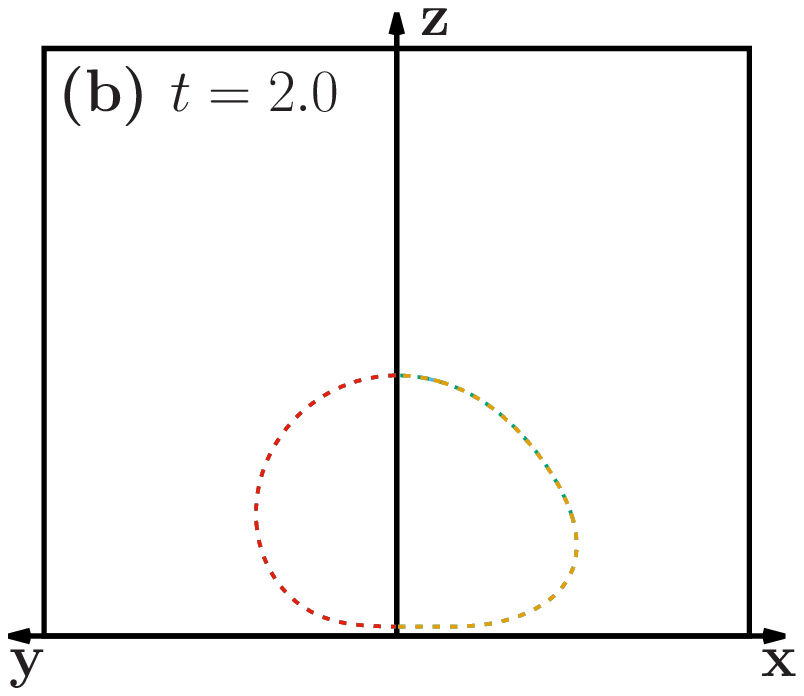}
	\includegraphics[trim= 20mm 1mm 20mm 1mm, clip, scale = 0.3, angle = 0]{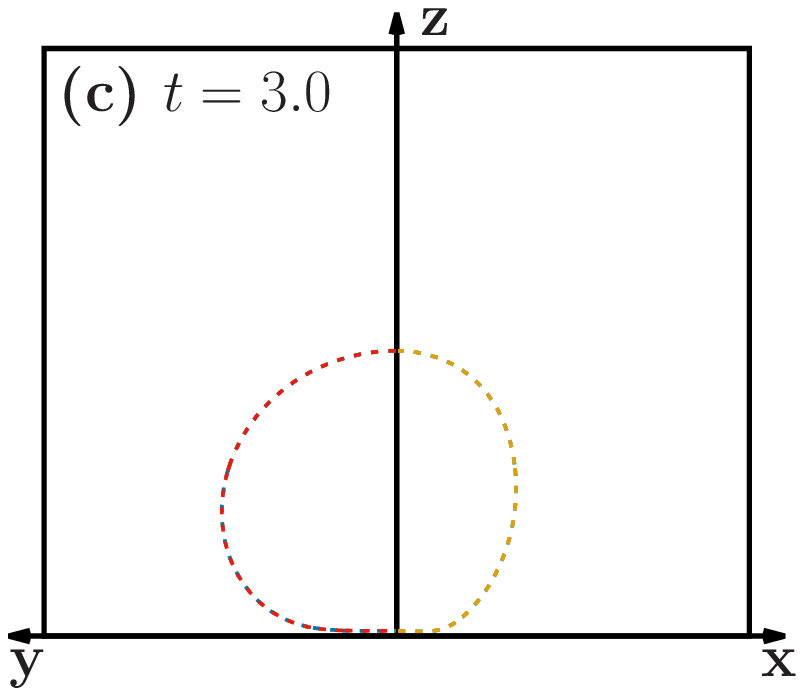}
	\includegraphics[trim= 20mm 1mm 20mm 1mm, clip, scale = 0.3, angle = 0]{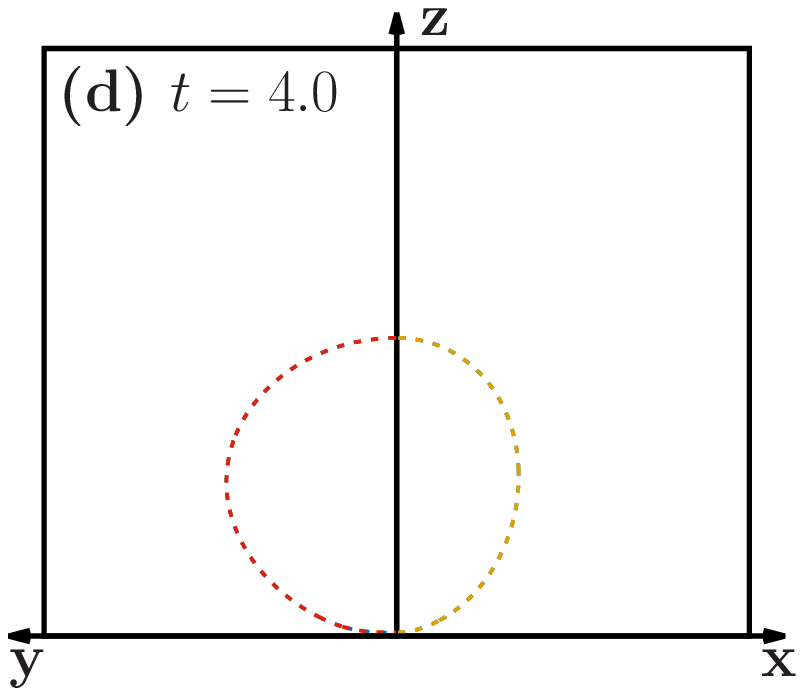}
	\includegraphics[trim= 20mm 1mm 20mm 1mm, clip, scale = 0.3, angle = 0]{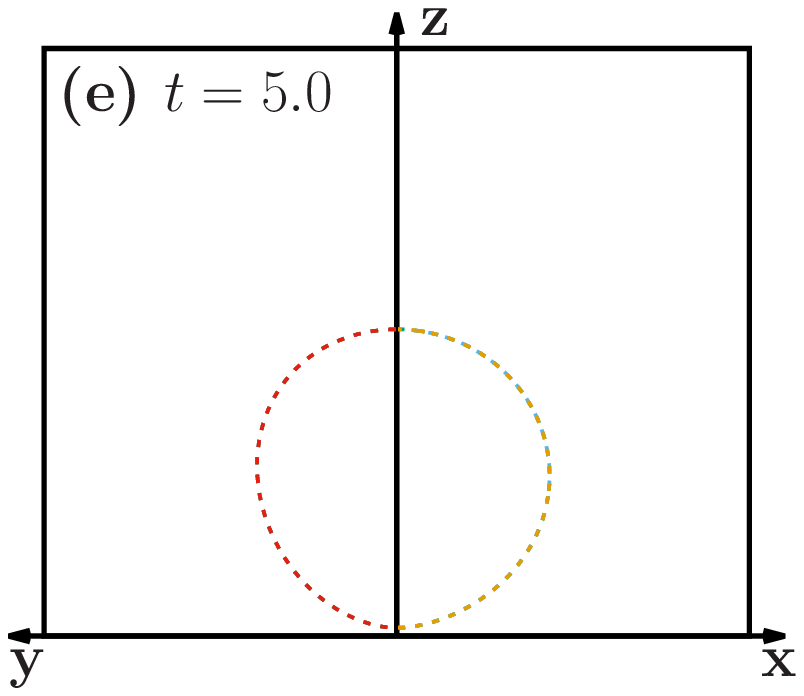}
	\includegraphics[trim= 20mm 1mm 20mm 1mm, clip, scale = 0.3, angle = 0]{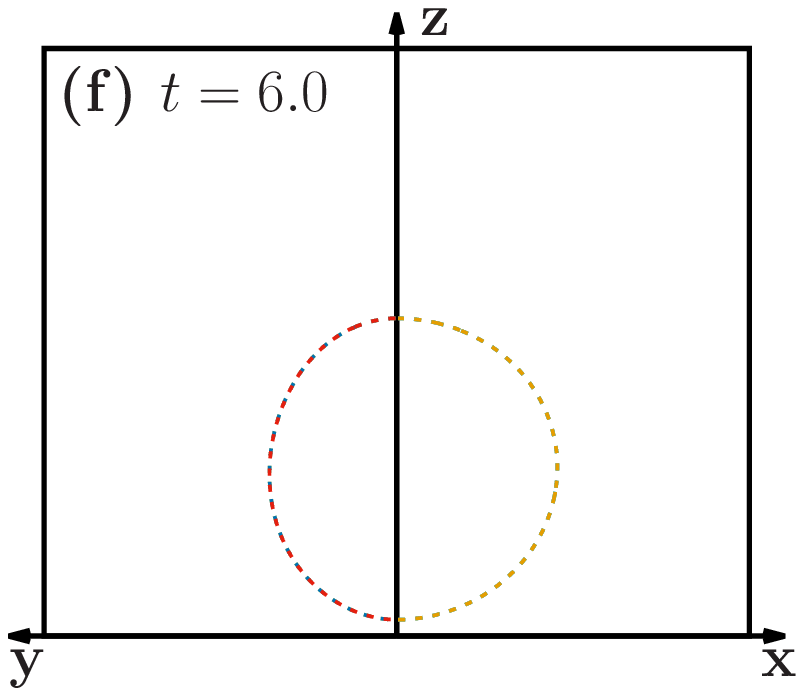}
	\caption{
		Snapshots of the drop shapes at $t= 1.0$, $2.0$, $3.0$, $4.0$, $5.0$ and $6.0$ (from left to right) at $Oh= 0.037$ for the coalescence induced drop jumping on a nonwetting wall by the present simulation 
		(upper row) and by the 3D MRT-LBM simulation (lower row).
		In each panel, the left part shows the $y-z$ plane at $x=0$
		and the right shows the $x-z$ plane at $y = 0$.
		For the upper row, 
		the two parts are arranged as in the 3D view.
	}
	\label{fig:jump-drop-t01-02-03-04-05-06}
\end{figure}

For this problem, the capillary-inertial velocity and time, 
$U_{ci} = \sqrt{\sigma / (\rho_{0} R)}$ and $T_{ci} = L_{c} / U_{ci}$, 
are used to scale the velocity and time quantities.
Two cases at $Oh=0.037$ ($Re=730.46$) and $Oh=0.119$ ($Re=70.616$) were simulated.
Figure \ref{fig:jump-drop-wdrop-evol} shows the evolutions of the centroid velocity of the drop in the $z-$direction $W_{drop}$ computed by the present method and by the 3D LBM (same as that in Section \ref{ssec:drop-dewetting}) using the same numerical parameters 
 ($N_{L} = 40$, $c = 10$, $Cn = 0.1$, $Pe = 8 \times 10^{3}$).
The grid size is $120 \times 120 \times 200$.
It can be found that the present results are in very good agreement with the LBM results for both $Oh=0.037$ and $0.119$.
In addition, several snapshots of the drop shape in two planes of symmetry are also shown in fig. \ref{fig:jump-drop-t01-02-03-04-05-06} to better illustrate the coalescence and jumping process at $Oh=0.037$.
One can see that the drop shapes by the SMAMEs are similar to those by the LBM at the selected times.
It was found that the drop jumped off the wall after some time when $Oh=0.037$ (e.g., see the snapshots at $t=6.0$ in fig. \ref{fig:jump-drop-t01-02-03-04-05-06})
and the drop always stayed on the wall when $Oh = 0.119$.
It is noted that the simulation time using the present SMAMEs is much shorter than that using the LBM.
For example, to run $40$ steps using four computational nodes on the same computer, 
the present method takes about $26$s whereas the LBM (D3Q19) takes about $70$s.
Here the LBM is parallelized by the MPI with the domain decomposed into four parts in the $x-$direction
and the SMAMEs is implemented in the AMReX framework~\cite{amrex19} with its default domain decomposition method (the solution of the CHE is similar in both solvers).
One can see that the present method  
is \emph{nearly three times faster} 
than the LBM
in the simulation of 3D two-phase flows.

\section{Concluding Remarks}\label{sec:conclusion}

To summarize, inspired by the MAMEs and LBM,
we have proposed a simplified numerical method to simulate incompressible viscous flows.
It was verified through a number of tests of both single- and two-phase flows
in 2D, axisymmetric and 3D geometries.
The results of all cases are as accurate as the LBM results and/or in good agreement with other reference results from analytical solutions or directly solving the incompressible NSEs.
At the same time, its implementation is much easier and the simulations using this new method cost much less memory and time than the corresponding LBM simulations. 
Some issues associated with the MAMEs, such as the use of intermediate variables and predictor-corrector step and the boundary conditions for additional derivatives, are no longer present in the new method.
Unlike the situation in the LBM, the inclusion of external forces is straightforward
since the macroscopic governing equations are handled directly.
For two-phase flows, a limitation of the present method is that it can only deal with flows with constant  viscosity and density (at most, with small density ratios).
In future, it will be further extended for flows with larger density and viscosity contrasts.
That may require more in-depth analyses of the LBEs for such problems.

\bf Acknowledgement \rm

This work is supported by the National Natural Science Foundation of China 
(NSFC, Grant No. 11972098).


\end{document}